\documentclass[11pt]{article}
\usepackage{cite}
\usepackage{mdframed}
\usepackage{epsfig,subfigure}
\usepackage{amssymb}
\usepackage[fleqn]{amsmath}
\usepackage{color}
\usepackage{arydshln}
\def\QED{\mbox{\rule[0pt]{1.5ex}{1.5ex}}}

\usepackage{graphicx}
\usepackage{color}
\usepackage[dvipsnames]{xcolor}

\definecolor{armygreen}{rgb}{0.29, 0.33, 0.13}

\usepackage{amssymb}
\usepackage{amsmath,amsfonts,amssymb}
\usepackage{verbatim}
\usepackage{stfloats}
\usepackage[bookmarks=false]{}
\usepackage{algorithm}
\usepackage{algcompatible}
\usepackage{tikz,pgfplots,tikzpeople}

\newtheorem{theorem}{Theorem}

\newtheorem{definition}{Definition}
\newtheorem{lemma}{Lemma}

\newtheorem{remark}{Remark}

\newtheorem{example}{Example}

\newcommand\blfootnote[1]{%
  \begingroup
  \renewcommand\thefootnote{}\footnote{#1}%
  \addtocounter{footnote}{-1}%
  \endgroup
}

\begin{document}
\date{}

\title{
On Extremal Rates of Secure Storage over Graphs
}
\author{\normalsize Zhou Li and Hua Sun \\
}

\maketitle

\blfootnote{
Zhou Li (email: zhouli@my.unt.edu) and Hua Sun (email: hua.sun@unt.edu) are with the Department of Electrical Engineering at the University of North Texas. }

\maketitle

\begin{abstract}
A secure storage code maps $K$ source symbols, each of $L_w$ bits, to $N$ coded symbols, each of $L_v$ bits, such that each coded symbol is stored in a node of a graph. Each edge of the graph is either associated with $D$ of the $K$ source symbols such that from the pair of nodes connected by the edge, we can decode the $D$ source symbols and learn no information about the remaining $K-D$ source symbols; or the edge is associated with no source symbols such that from the pair of nodes connected by the edge, nothing about the $K$ source symbols is revealed. The ratio $L_w/L_v$ is called the symbol rate of a secure storage code and the highest possible symbol rate is called the capacity.

We characterize all graphs over which the capacity of a secure storage code is equal to $1$, when $D = 1$. This result is generalized to $D> 1$, i.e., we characterize all graphs over which the capacity of a secure storage code is equal to $1/D$ under a mild condition that for any node, the source symbols associated with each of its connected edges do not include a common element. Further, we characterize all graphs over which the capacity of a secure storage code is equal to $2/D$.
\end{abstract}

\newpage

\allowdisplaybreaks
\section{Introduction}
Modern datasets are usually massive and stored in a distributed manner. Providing flexible accessibility and security control over a variety of network topologies with limited storage budget is a challenging task. Motivated by such secure storage tasks, in this work we model a distributed storage system and its data access structure using a graph and aim to find storage efficient codes that satisfy the accessibility and security constraints specified by the graph.

A secure storage code is a mapping from $K$ source symbols, $W_1, \cdots, W_K$, each of $L_w$ bits, to $N$ coded symbols, $V_1, \cdots, V_N$, each of $L_v$ bits. Each coded symbol is stored in a node of a graph $G$, so the node set of the graph is $\mathcal{V} = \{V_1, \cdots, V_N\}$. Note that we use $V_n$ to denote both a coded symbol and a node as they have a one-to-one mapping. The data accessibility and security constraints are given through the edges of the graph. An edge $\{V_i, V_j\}$ of the graph $G$ is associated either with $D$ of the $K$ source symbols or no source symbols. In the former case, the requirement is that from $(V_i, V_j)$, we can decode the $D$ source symbols and learn nothing about the remaining $K-D$ source symbols; in the latter case, $(V_i, V_j)$ must be independent of the $K$ source symbols such that no information is leaked. An example of a secure storage problem over a graph is given in Fig.~\ref{fig:prob}. The storage efficiency of a secure storage code is measured by its symbol rate, defined as $L_w/L_v$, i.e., out of the ($L_v$) bits of each coded symbol, how many bits ($L_w$) of each source symbol can be securely stored. Our objective is to characterize for a given graph $G$, the highest possible symbol rate, termed the capacity $C = \sup L_w/L_v$, of a secure storage code.

\vspace{0.1in}
\tikzset{
    photon/.style={decorate, decoration={snake,
      amplitude = 0.5mm,
      segment length = 3mm}, draw=violet},
    electron/.style={draw=blue, postaction={decorate},
        decoration={markings,mark=at position .55 with {\arrow[draw=blue]{>}}}},
    gluon/.style={decorate, draw=blue,
        decoration={coil,amplitude=1.5pt, segment length=4pt}} 
}
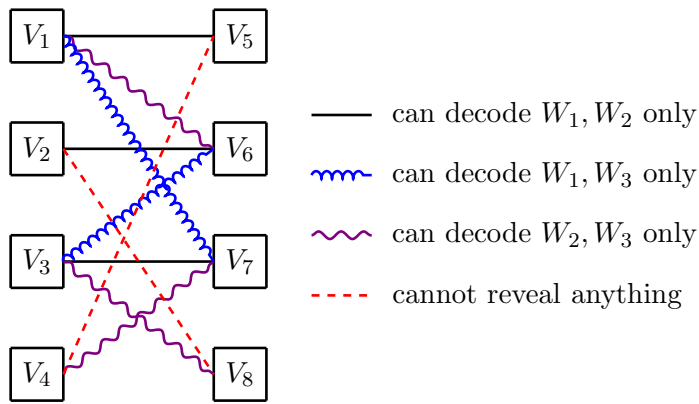
\begin{figure}[H]
\begin{center}
\begin{tikzpicture}
\filldraw (1.35,-1.15) node {$V_4$};
\filldraw (1.35,0.35) node {$V_3$};
\filldraw (1.35,1.85) node {$V_2$};
\filldraw (1.35,3.35) node {$V_1$};
\filldraw (4.05,-1.15) node {$V_8$};
\filldraw (4.05,0.35) node {$V_7$};
\filldraw (4.05,1.85) node {$V_6$};
\filldraw (4.05,3.35) node {$V_5$};

\draw[black, line width=1] (1,-1.5) -- (1,-0.8);
\draw[black, line width=1] (1,-1.5) -- (1.7,-1.5);
\draw[black, line width=1] (1,-0.8) -- (1.7,-0.8);
\draw[black, line width=1] (1.7,-1.5) -- (1.7,-0.8);

\draw[black, line width=1] (3.7,-1.5) -- (3.7,-0.8);
\draw[black, line width=1] (3.7,-1.5) -- (4.4,-1.5);
\draw[black, line width=1] (3.7,-0.8) -- (4.4,-0.8);
\draw[black, line width=1] (4.4,-1.5) -- (4.4,-0.8);

\draw[black, line width=1] (1,0) -- (1,0.7);
\draw[black, line width=1] (1,0) -- (1.7,0);
\draw[black, line width=1] (1,0.7) -- (1.7,0.7);
\draw[black, line width=1] (1.7,0) -- (1.7,0.7);

\draw[black, line width=1] (3.7,0) -- (3.7,0.7);
\draw[black, line width=1] (3.7,0) -- (4.4,0);
\draw[black, line width=1] (3.7,0.7) -- (4.4,0.7);
\draw[black, line width=1] (4.4,0) -- (4.4,0.7);

\draw[black, line width=1] (1,1.5) -- (1,2.2);
\draw[black, line width=1] (1,1.5) -- (1.7,1.5);
\draw[black, line width=1] (1,2.2) -- (1.7,2.2);
\draw[black, line width=1] (1.7,1.5) -- (1.7,2.2);

\draw[black, line width=1] (3.7,1.5) -- (3.7,2.2);
\draw[black, line width=1] (3.7,1.5) -- (4.4,1.5);
\draw[black, line width=1] (3.7,2.2) -- (4.4,2.2);
\draw[black, line width=1] (4.4,1.5) -- (4.4,2.2);

\draw[black, line width=1] (1,3) -- (1,3.7);
\draw[black, line width=1] (1,3) -- (1.7,3);
\draw[black, line width=1] (1,3.7) -- (1.7,3.7);
\draw[black, line width=1] (1.7,3) -- (1.7,3.7);

\draw[black, line width=1] (3.7,3) -- (3.7,3.7);
\draw[black, line width=1] (3.7,3) -- (4.4,3);
\draw[black, line width=1] (3.7,3.7) -- (4.4,3.7);
\draw[black, line width=1] (4.4,3) -- (4.4,3.7);

\draw[black, line width=1] (1.7,3.35) -- (3.7,3.35);
\draw[black, line width=1] (1.7,1.85) -- (3.7,1.85);
\draw[black, line width=1] (1.7,0.35) -- (3.7,0.35);

\draw[green, photon, line width=1] (1.7,3.35) -- (3.7,1.85);
\draw[green, photon, line width=1] (1.7,0.35) -- (3.7,-1.15);
\draw[green, photon, line width=1] (1.7,-1.15) -- (3.7,0.35);

\draw[black, gluon, line width=1] (1.7,3.35) -- (3.7,0.35);
\draw[black, gluon, line width=1] (1.7,0.35) -- (3.7,1.85);

\draw[red, dashed, line width = 1] (1.7,1.85) -- (3.7,-1.15);
\draw[red, dashed, line width = 1] (1.7,-1.15) -- (3.7,3.35);

\draw[black, line width=1] (5, 2.3) -- (5.8,2.3);
\draw[black, gluon, line width=1](5, 1.5) -- (5.8, 1.5);
\draw[black, photon, line width=1](5, 0.7) -- (5.8, 0.7);
\draw[red, dashed, line width=1](5, -0.1) -- (5.8,-0.1);

\filldraw (8.1,2.3) node {can decode $W_1,W_2$ only};
\filldraw (8.1,1.5) node {can decode $W_1,W_3$ only};
\filldraw (8.1,0.7) node {can decode $W_2,W_3$ only};
\filldraw (8.0,-0.1) node {cannot reveal anything};

\end{tikzpicture}

\end{center}
\vspace{-0.1in}
  \caption{\small An example graph of a secure storage problem with $K=3$ source symbols and $N=8$ coded symbols, whose capacity turns out to be $1/2$ (refer to Theorem \ref{thm:d2}. See Fig.~\ref{fig:ex3} for a code construction).
  }
  \label{fig:prob}
\end{figure}

While this work is presented in a storage system context, the problem of secure storage has intimate relations to a few communication network contexts. First, when the graph $G$ is bipartite (e.g., Fig.~\ref{fig:prob}), the secure storage problem can be viewed a generalization of the conditional disclosure of secrets (CDS) problem \cite{SymPIR, Applebaum_Arkis_Raykov_Vasudevan,Li_Sun_CDS, Li_Sun_linearCDS, Wang_Ulukus_CDMS}. To see this, we view the nodes on one side (e.g., $V_1, V_2, V_3, V_4$ in Fig.~\ref{fig:prob}) as the transmit signal sent by Alice and view the nodes on the other side (e.g., $V_5, V_6, V_7, V_8$ in Fig.~\ref{fig:prob}) as the transmit signal sent by Bob. If and only if the signal indices (node indices) satisfy some function (i.e., the type of the edge corresponds to some source symbols), Carol who receives both signals can recover the corresponding secrets. Compared to the classic CDS problem where there is only one secret (source symbol) to disclose, the secure storage problem generalizes to include multiple secrets \cite{Wang_Ulukus_CDMS}; further, an arbitrary subset of all secrets can be conditionally disclosed. Second, the secure storage problem can be interpreted as a secure network coding problem \cite{Ahlswede_Cai_etal, secure_nc, Cai_Chan} over a class of combination networks. While previous work mainly considers multicast or special message structures (e.g., nested) over combination networks (often with no security) \cite{combination_network, Maheshwar_Li_Li, bidokhti2016capacity, Salimi_Liu_Cui}, our focus in this work is on the interplay between data access and security pattern modelled by graphs other than the network topology graph.

\subsection*{Main Result and Technique}
Characterizing the capacity of secure storage codes appears to be a formidable task, mainly due to the fact that the constraint graph $G$ can be arbitrary. Different classes of graphs (or hypergraphs) $G$ can be used to model various well-known network information theory problems, such as index coding \cite{Yossef_Birk_Jayram_Kol_Trans} and coded caching \cite{Maddah_Ali_Niesen}, as noticed in \cite{Sahraei_Gastpar}, which considers a similar graph storage problem with no security constraints (but the coded symbols may have different sizes). As a result, allowing arbitrary $G$ will involve well-established hard capacity questions.

Instead of fixing a graph $G$ first and then pursuing the capacity, the perspective we take in this work is to focus on {\em extremal} rate values and the associated {\em extremal} graphs whose secure storage capacity values are extremal. A natural starting point is the setting of $D = 1$ and $C=1$, where it is easily seen that the capacity of secure storage code cannot exceed $1$, i.e., the size of each coded symbol must be at least the size of each source symbol, as long as there exist security constraints. Our first main result is a full characterization of all such extremal graphs whose capacity is $C=1$ (refer to Theorem \ref{thm:d1}), i.e., if a graph belongs to this class, we construct a secure storage code that achieves the highest possible symbol rate of $1$ and otherwise if a graph does not belong to this class, the symbol rate of any secure storage code must be strictly smaller than $1$. The key to this extremal rate characterization result is an alignment view of the space of the source symbols, the noise symbols (required to ensure information theoretic security), and the coded symbols. Such an alignment view is introduced in \cite{Li_Sun_CDS}, where all extremal graphs with $C=1$ are found with $K=1$ source symbol. This work generalizes this result to an arbitrary number of source symbols. While only noise alignment and signal (coded symbol) alignment are needed in \cite{Li_Sun_CDS} as there is only $K=1$ source symbol, here we further need interference alignment to take care of other undesired source symbols as $K > 1$. Interestingly, a decomposition based approach turns out to be effective, i.e., we first separately design a secure storage code for each source symbol and then combine each separate code to produce a joint code that works for all source symbols.

Our second main result is a generalization of Theorem \ref{thm:d1} from $D=1$ to any $D > 1$, but under an additional condition to ensure that each coded symbol must be fully covered by noise symbols (then $C\leq 1/D$, equivalently, $L_v \geq D \times L_w$). Under such a condition, we characterize all extremal graphs whose capacity is $C=1/D$ (refer to Theorem \ref{thm:d2}). Compared to Theorem \ref{thm:d1} where each edge may recover $D=1$ source symbol, Theorem \ref{thm:d2} considers the case where each edge may recover $D>1$ source symbols and this introduces some technical difficulty. While the same decomposition based approach continues to apply, ensuring the simultaneous recovery of multiple source symbols is more involved. As a consequence of such difficulty, the code construction in Theorem \ref{thm:d1} is explicit while in Theorem \ref{thm:d2} we are only able to provide an existence proof that relies on randomized code constructions.

Finally, noting that there exist graphs whose secure storage code rates are strictly larger than $C = 1/D$, we study the extremal rate of $2/D$, which is the highest possible symbol rate among all graphs. Here any pair of nodes connected by an edge have a total storage size of $2L_v = D\times L_w$, i.e., all storage space is occupied by the desired $D$ source symbols. This extremal rate of $2/D$ places very strict constraints on the graph $G$. Our third main result is a full characterization of all extremal graphs $G$ whose capacity is $C=2/D$ (refer to Theorem \ref{thm:2d}). Notably, linear coding (storing linear combinations of different source symbols) is necessary to achieve the capacity of $2/D$, i.e., storing the source symbols directly is not sufficient.

\section{Problem Statement and Definitions}\label{sec:model}
Consider $K$ independent source symbols $W_1, \cdots, W_K$, each of $L_w$ bits.
\begin{eqnarray}
&& H(W_1, \cdots, W_K) = H(W_1) + \cdots + H(W_K), \notag\\
&& L_w = H(W_1) = \cdots = H(W_K). \label{h1}
\end{eqnarray}

Consider $N$ coded symbols $V_1, \cdots, V_N$, each of $L_v$ bits. Note that $L_w, L_v$ are not necessarily integer values. For example, if $W_k$ are uniformly random $\mathbb{F}_3$ symbols, then $L_w = \log_2 3$ bits. Furthermore, since we are interested in their relative size (see (\ref{rate})), $L_w, L_v$ are allowed to take arbitrarily large values. 

The constraints on the coded symbols are specified by a graph $G = (\mathcal{V}, \mathcal{E})$, where the node\footnote{Note that we abuse the notation by using $V_n$ to denote both a coded symbol and a node of the graph, for the sake of simplicity. The context will make its meaning clear.} set $\mathcal{V} = \{V_1, \cdots, V_N\}$ and the edge set $\mathcal{E}$ is a set of unordered pairs from $\mathcal{V}$. Each edge $\{V_i, V_j\} \in \mathcal{E}$ is associated with a subset $\mathcal{D}$ of $\{1,2,\cdots,K\}\triangleq [K]$, which either has $D$ elements or is an empty set, i.e., $|\mathcal{D}| = D$ or $\mathcal{D} = \emptyset$. The edge association is described by a function $t: t(\{V_i, V_j\}) = \mathcal{D}$. For each edge $\{V_i,V_j\}$, it is required that from $(V_i, V_j)$, we can decode and only decode the messages $\left(W_k \right)_{k \in \mathcal{D}}$.
That is, $\forall \{V_i, V_j\} \in \mathcal{E}$ such that $t(\{V_i, V_j\}) = \mathcal{D}$, we have
\begin{eqnarray}
(\mbox{Correctness}) && H\left( \left(W_k\right)_{k \in \mathcal{D}} \mid V_i, V_j \right) = 0, \label{dec} \\
(\mbox{Security}) && I\left(V_i, V_j; \left(W_k\right)_{k \in [K]\backslash\mathcal{D}}| \left(W_k\right)_{k \in \mathcal{D}} \right) = 0 \label{sec} 
\end{eqnarray}
where for two sets $\mathcal{A}, \mathcal{B}$, $\mathcal{A}\backslash\mathcal{B}$ denotes 
the set of elements that belong to $\mathcal{A}$ but not to $\mathcal{B}$. An isolated node $V$, i.e., a node connected to no edges, is trivial as it has no constraint. Without loss of generality, we assume that any graph $G$ considered in this work contains no isolated nodes.

A mapping from the source symbols $W_1, \cdots, W_K$ to the coded symbols $V_1, \cdots, V_N$ that satisfies the correctness and security constraints (\ref{dec}), (\ref{sec}) specified by a graph $G = (\mathcal{V}, \mathcal{E})$ is called a secure storage code. The (achievable) symbol rate of a secure storage code is defined as
\begin{eqnarray}
R \triangleq \frac{L_w}{L_v} \label{rate}
\end{eqnarray}
and the supremum of symbol rates is called the capacity, $C$. Note that supremum includes limits, so $R = \lim_{L_w\rightarrow \infty} L_w/L_v$ is also (asymptotically) achievable.

\subsection{Graph Definitions}
To facilitate the presentation of our results, we introduce some graph definitions in this section.

For a graph $G = (\mathcal{V}, \mathcal{E})$, we wish to separately consider each source symbol $W_k$ and see if each edge is associated with $W_k$ (i.e., can recover $W_k$). This leads us to the definition of $G^{[k]}$.
\begin{definition}[Characteristic Graph $G^{[k]}$ of $W_k$] \label{def:cha}
For a graph $G = (\mathcal{V}, \mathcal{E})$, define $\forall k \in [K]$
\begin{eqnarray}
&& G^{[k]} = (\mathcal{V}^{[k]}, \mathcal{E}^{[k]}) ~\mbox{such that}~  \mathcal{V}^{[k]} = \{V^{[k]}_1, \cdots, V^{[k]}_{N}\}, \notag\\
&& \{V^{[k]}_i, V^{[k]}_j \} \in \mathcal{E}^{[k]} ~\mbox{if and only if}~  \{V_i, V_j\} \in \mathcal{E}, \notag\\
&& t^{[k]}(\{V^{[k]}_i, V^{[k]}_j \}) =
\left\{\begin{array}{cl}
\{k\} & ~\mbox{if}~  k \in t(\{V_i, V_j \}) \\
\emptyset & ~\mbox{else if}~ k \notin t(\{V_i, V_j \})
\end{array}
\right..
\end{eqnarray}
\end{definition}

Fig.~\ref{fig:prob1} shows an example of $G$ and its $G^{[1]}$ of $W_1$.

\vspace{0.1in}
\tikzset{
    photon/.style={decorate, decoration={snake,
      amplitude = 0.5mm,
      segment length = 3mm}, draw=violet},
    electron/.style={draw=blue, postaction={decorate},
        decoration={markings,mark=at position .55 with {\arrow[draw=blue]{>}}}},
    gluon/.style={decorate, draw=blue,
        decoration={coil,amplitude=1.5pt, segment length=4pt}} 
}
\begin{figure}[H]
\begin{center}
\subfigure[]{
\begin{tikzpicture}
\filldraw (1.35,-1.15) node {$V_4$};
\filldraw (1.35,0.35) node {$V_3$};
\filldraw (1.35,1.85) node {$V_2$};
\filldraw (1.35,3.35) node {$V_1$};
\filldraw (4.05,-1.15) node {$V_8$};
\filldraw (4.05,0.35) node {$V_7$};
\filldraw (4.05,1.85) node {$V_6$};
\filldraw (4.05,3.35) node {$V_5$};

\draw[black, line width=1] (1,-1.5) -- (1,-0.8);
\draw[black, line width=1] (1,-1.5) -- (1.7,-1.5);
\draw[black, line width=1] (1,-0.8) -- (1.7,-0.8);
\draw[black, line width=1] (1.7,-1.5) -- (1.7,-0.8);

\draw[black, line width=1] (3.7,-1.5) -- (3.7,-0.8);
\draw[black, line width=1] (3.7,-1.5) -- (4.4,-1.5);
\draw[black, line width=1] (3.7,-0.8) -- (4.4,-0.8);
\draw[black, line width=1] (4.4,-1.5) -- (4.4,-0.8);

\draw[black, line width=1] (1,0) -- (1,0.7);
\draw[black, line width=1] (1,0) -- (1.7,0);
\draw[black, line width=1] (1,0.7) -- (1.7,0.7);
\draw[black, line width=1] (1.7,0) -- (1.7,0.7);

\draw[black, line width=1] (3.7,0) -- (3.7,0.7);
\draw[black, line width=1] (3.7,0) -- (4.4,0);
\draw[black, line width=1] (3.7,0.7) -- (4.4,0.7);
\draw[black, line width=1] (4.4,0) -- (4.4,0.7);

\draw[black, line width=1] (1,1.5) -- (1,2.2);
\draw[black, line width=1] (1,1.5) -- (1.7,1.5);
\draw[black, line width=1] (1,2.2) -- (1.7,2.2);
\draw[black, line width=1] (1.7,1.5) -- (1.7,2.2);

\draw[black, line width=1] (3.7,1.5) -- (3.7,2.2);
\draw[black, line width=1] (3.7,1.5) -- (4.4,1.5);
\draw[black, line width=1] (3.7,2.2) -- (4.4,2.2);
\draw[black, line width=1] (4.4,1.5) -- (4.4,2.2);

\draw[black, line width=1] (1,3) -- (1,3.7);
\draw[black, line width=1] (1,3) -- (1.7,3);
\draw[black, line width=1] (1,3.7) -- (1.7,3.7);
\draw[black, line width=1] (1.7,3) -- (1.7,3.7);

\draw[black, line width=1] (3.7,3) -- (3.7,3.7);
\draw[black, line width=1] (3.7,3) -- (4.4,3);
\draw[black, line width=1] (3.7,3.7) -- (4.4,3.7);
\draw[black, line width=1] (4.4,3) -- (4.4,3.7);

\draw[black, line width=1] (1.7,3.35) -- (3.7,3.35);
\draw[black, line width=1] (1.7,1.85) -- (3.7,1.85);
\draw[black, line width=1] (1.7,0.35) -- (3.7,0.35);

\draw[green, photon, line width=1] (1.7,3.35) -- (3.7,1.85);
\draw[green, photon, line width=1] (1.7,0.35) -- (3.7,-1.15);
\draw[green, photon, line width=1] (1.7,-1.15) -- (3.7,0.35);

\draw[black, gluon, line width=1] (1.7,3.35) -- (3.7,0.35);
\draw[black, gluon, line width=1] (1.7,0.35) -- (3.7,1.85);

\draw[red, dashed, line width = 1] (1.7,1.85) -- (3.7,-1.15);
\draw[red, dashed, line width = 1] (1.7,-1.15) -- (3.7,3.35);

\draw[black, line width=1] (5, 2.3) -- (5.8,2.3);
\draw[black, gluon, line width=1](5, 1.5) -- (5.8, 1.5);
\draw[black, photon, line width=1](5, 0.7) -- (5.8, 0.7);
\draw[red, dashed, line width=1](5, -0.1) -- (5.8,-0.1);

\filldraw (6.5,3) node {$t$};
\filldraw (6.5,2.3) node {$\{1,2\}$};
\filldraw (6.5,1.5) node {$\{1,3\}$};
\filldraw (6.5,0.7) node {$\{2,3\}$};
\filldraw (6.5,-0.1) node {$\emptyset$};

\end{tikzpicture}
}
\hspace{0.4in}
\subfigure[]{ 
\begin{tikzpicture}
\filldraw (1.35,-1.15) node {$V^{[1]}_4$};
\filldraw (1.35,0.35) node {$V^{[1]}_3$};
\filldraw (1.35,1.85) node {$V^{[1]}_2$};
\filldraw (1.35,3.35) node {$V^{[1]}_1$};
\filldraw (4.05,-1.15) node {$V^{[1]}_8$};
\filldraw (4.05,0.35) node {$V^{[1]}_7$};
\filldraw (4.05,1.85) node {$V^{[1]}_6$};
\filldraw (4.05,3.35) node {$V^{[1]}_5$};

\draw[black, line width=1] (1,-1.5) -- (1,-0.8);
\draw[black, line width=1] (1,-1.5) -- (1.7,-1.5);
\draw[black, line width=1] (1,-0.8) -- (1.7,-0.8);
\draw[black, line width=1] (1.7,-1.5) -- (1.7,-0.8);

\draw[black, line width=1] (3.7,-1.5) -- (3.7,-0.8);
\draw[black, line width=1] (3.7,-1.5) -- (4.4,-1.5);
\draw[black, line width=1] (3.7,-0.8) -- (4.4,-0.8);
\draw[black, line width=1] (4.4,-1.5) -- (4.4,-0.8);

\draw[black, line width=1] (1,0) -- (1,0.7);
\draw[black, line width=1] (1,0) -- (1.7,0);
\draw[black, line width=1] (1,0.7) -- (1.7,0.7);
\draw[black, line width=1] (1.7,0) -- (1.7,0.7);

\draw[black, line width=1] (3.7,0) -- (3.7,0.7);
\draw[black, line width=1] (3.7,0) -- (4.4,0);
\draw[black, line width=1] (3.7,0.7) -- (4.4,0.7);
\draw[black, line width=1] (4.4,0) -- (4.4,0.7);

\draw[black, line width=1] (1,1.5) -- (1,2.2);
\draw[black, line width=1] (1,1.5) -- (1.7,1.5);
\draw[black, line width=1] (1,2.2) -- (1.7,2.2);
\draw[black, line width=1] (1.7,1.5) -- (1.7,2.2);

\draw[black, line width=1] (3.7,1.5) -- (3.7,2.2);
\draw[black, line width=1] (3.7,1.5) -- (4.4,1.5);
\draw[black, line width=1] (3.7,2.2) -- (4.4,2.2);
\draw[black, line width=1] (4.4,1.5) -- (4.4,2.2);

\draw[black, line width=1] (1,3) -- (1,3.7);
\draw[black, line width=1] (1,3) -- (1.7,3);
\draw[black, line width=1] (1,3.7) -- (1.7,3.7);
\draw[black, line width=1] (1.7,3) -- (1.7,3.7);

\draw[black, line width=1] (3.7,3) -- (3.7,3.7);
\draw[black, line width=1] (3.7,3) -- (4.4,3);
\draw[black, line width=1] (3.7,3.7) -- (4.4,3.7);
\draw[black, line width=1] (4.4,3) -- (4.4,3.7);

\draw[black, line width=1] (1.7,3.35) -- (3.7,3.35);
\draw[black, line width=1] (1.7,1.85) -- (3.7,1.85);
\draw[black, line width=1] (1.7,0.35) -- (3.7,0.35);

\draw[red, dashed, line width = 1] (1.7,3.35) -- (3.7,1.85);
\draw[red, dashed, line width = 1] (1.7,0.35) -- (3.7,-1.15);
\draw[red, dashed, line width = 1] (1.7,-1.15) -- (3.7,0.35);

\draw[black, line width=1] (1.7,3.35) -- (3.7,0.35);
\draw[black, line width=1] (1.7,0.35) -- (3.7,1.85);

\draw[red, dashed, line width = 1] (1.7,1.85) -- (3.7,-1.15);
\draw[red, dashed, line width = 1] (1.7,-1.15) -- (3.7,3.35);

\draw[black, line width=1] (5,1.5) -- (5.8,1.5);
\draw[red, dashed, line width=1](5, 0.7) -- (5.8,0.7);

\filldraw (6.3,2.3) node {$t^{[1]}$};
\filldraw (6.3,1.5) node {$\{1\}$};
\filldraw (6.3,0.7) node {$\emptyset$};

\end{tikzpicture}
 }
\end{center}
\vspace{-0.2in}
  \caption{\small (a) An example graph $G$ and (b) its characteristic graph $G^{[1]}$ of $W_1$.
  }
  \label{fig:prob1}
\end{figure}
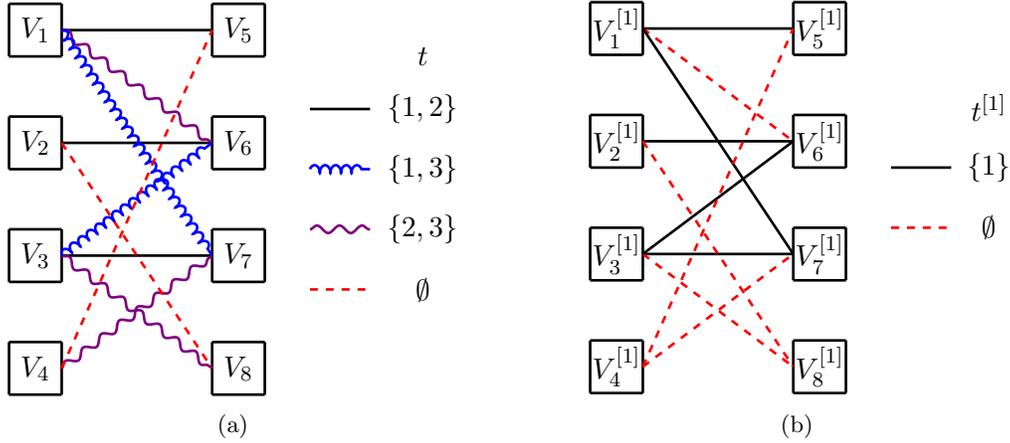

For a node $V$ of a graph $G = (\mathcal{V}, \mathcal{E})$, the common elements of the source symbols associated with each of its connected edges are relevant in stating our results, then we make them explicit in the following definition.

\begin{definition}[Common Sources $\mathcal{C}(V)$]
Consider a node $V\in\mathcal{V}$ of a graph $G = (\mathcal{V}, \mathcal{E})$, define
\begin{eqnarray}
\mathcal{C}(V) = \bigcap_{i: \{V, V_i\} \in \mathcal{E}}~ t(\{V, V_i\}).
\end{eqnarray}
\end{definition}

For example, consider node $V_1$ in Fig.~\ref{fig:prob1}, $\mathcal{C}(V_1) = \{1,2\} \cap \{2,3\} \cap \{1,3\} = \emptyset$.
\vspace{0.05in}

For an edge of a graph $G = (\mathcal{V}, \mathcal{E})$, it is important if the edge is associated with some source symbol or no source symbol. Depending on this, an edge is called either qualified or unqualified and we have similar definitions for paths and components.

\begin{definition}[Qualified/Unqualified Edge/Path/Component]
Consider a graph $G = (\mathcal{V}, \mathcal{E})$. An edge $E \in \mathcal{E}$ is called qualified if $t(E) \neq \emptyset$ and unqualified if $t(E) = \emptyset$. A sequence of connecting qualified/unqualified edges is called a qualified/unqualified path. A qualified edge that connects two nodes in an unqualified path is said to be internal. A qualified/unqualified component is a maximal induced subgraph of $G$ wherein any two nodes are connected by a qualified/unqualified path.
\end{definition}

Note that the above definition applies to both $G$ and $G^{[k]}$. For example, in Fig.~\ref{fig:prob1}, $\{V_1, V_5\}$ is a qualified edge, $\{V_1^{[1]}, V_6^{[1]}\}$ is an unqualified edge, $\left(\{V_5^{[1]}, V_4^{[1]}\}, \{V_4^{[1]}, V_7^{[1]}\}\right)$ is an unqualified path, $G$ is a qualified component, and $G^{[1]}$ contains no internal qualified edges.
\vspace{0.05in}

Finally, a node $V$ of a graph $G = (\mathcal{V}, \mathcal{E})$ whose all connected edges are associated with the same set of source symbols, is degenerate (because all constraints of $V$ can be satisfied by storing the same set of source symbols in $V$). It is convenient to remove all degenerate nodes when the results are presented and we have the following definition.
\begin{definition}[Non-degenerate Subgraph $\widetilde{G}$ of $G$]
For a graph $G = (\mathcal{V}, \mathcal{E})$, denote the set of degenerate nodes by $\mathcal{V}_d$, i.e., 
\begin{eqnarray}
\mathcal{V}_d \triangleq \bigcup ~\big\{V\in\mathcal{V} \mid t(\{V, V_i\}) = \mathcal{C}(V), ~ \forall \{V, V_i\} \in \mathcal{E} \big\}.
\end{eqnarray}
The subgraph of $G$ induced by the non-degenerate nodes $\mathcal{V}\backslash\mathcal{V}_d$ is defined as $\widetilde{G}$, i.e., $\widetilde{G} \triangleq G[\mathcal{V}\backslash\mathcal{V}_d]$.
\end{definition}

\section{Results}
In this section, we present our main results along with illustrative examples and observations.

\subsection{$D=1$ and Extremal Graphs with $C=1$}
We start with the setting of $D = 1$. All extremal graphs whose secure storage capacity is $C=1$ are characterized in the following theorem.

\begin{theorem}\label{thm:d1}
The capacity of a secure storage code over a graph $G$ with $D=1$ is $C=1$ if and only if the non-degenerate subgraph $\widetilde{G}$ of $G$ is not empty and for every qualified component $Q$ of $\widetilde{G}$, the characteristic graph $Q^{[k]}$ of each coded symbol $W_k, k \in [K]$ contains no internal qualified edge.
\end{theorem}

\begin{remark}
When the non-degenerate subgraph $\widetilde{G}$ of $G$ is empty, each node of $G$ is connected to edges that are associated with the same set of source symbols. If there exists a qualified edge in $G$, the secure storage capacity is $2$ (this case will be covered in Theorem \ref{thm:2d}).
\end{remark}

The proof of Theorem \ref{thm:d1} is presented in Section \ref{sec:thm1}.
Here to illustrate the idea, we give two examples. The first example (see Fig.~\ref{fig:ex1}) is used to explain the `if' part, i.e., the graph $G$ satisfies the condition in Theorem \ref{thm:d1} and the secure storage capacity is $C=1$.

\tikzset{
    photon/.style={decorate, decoration={snake,
      amplitude = 0.5mm,
      segment length = 3mm}, draw=green},
    electron/.style={draw=blue, postaction={decorate},
        decoration={markings,mark=at position .55 with {\arrow[draw=blue]{>}}}},
    gluon/.style={decorate, draw=blue,
        decoration={coil,amplitude=1.5pt, segment length=4pt}} 
}
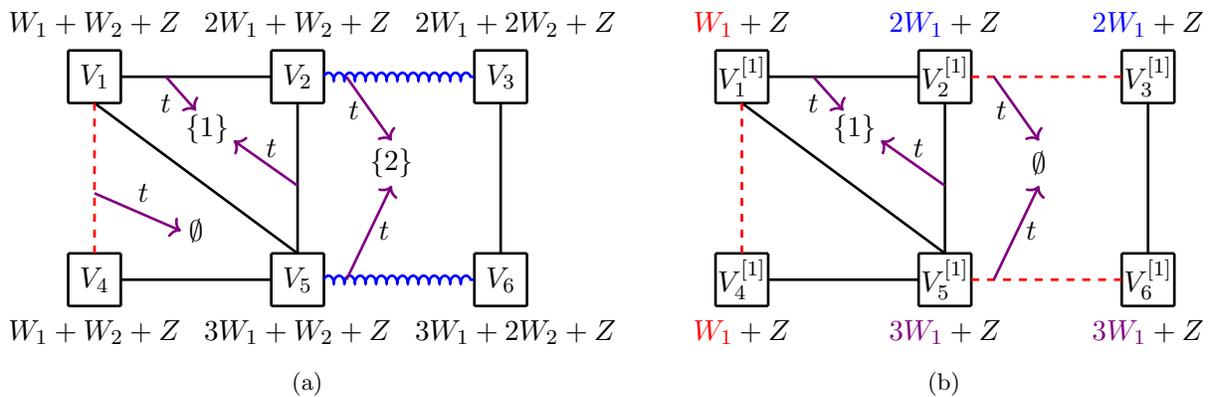
\begin{figure}[H]
\begin{center}
\subfigure[]{
\begin{tikzpicture}
\filldraw (0.35,0.35) node {$V_4$};
\filldraw (5.75,3.05) node {$V_3$};
\filldraw (3.05,3.05) node {$V_2$};
\filldraw (0.35,3.05) node {$V_1$};
\filldraw (5.75,0.35) node {$V_6$};
\filldraw (3.05,0.35) node {$V_5$};

\filldraw (0.35,3.75) node {$W_1+W_2+Z$};
\filldraw (0.35,-0.35) node {$W_1+W_2+Z$};

\filldraw (5.95,3.75) node {$2W_1+2W_2+Z$};
\filldraw (3.05,3.75) node {$2W_1+W_2+Z$};

\filldraw (5.95,-0.35) node {$3W_1+2W_2+Z$};
\filldraw (3.05,-0.35) node {$3W_1+W_2+Z$};

\draw[black, line width=1] (0,0) -- (0,0.7);
\draw[black, line width=1] (0,0) -- (0.7,0);
\draw[black, line width=1] (0,0.7) -- (0.7,0.7);
\draw[black, line width=1] (0.7,0) -- (0.7,0.7);

\draw[black, line width=1] (2.7,0) -- (2.7,0.7);
\draw[black, line width=1] (2.7,0) -- (3.4,0);
\draw[black, line width=1] (2.7,0.7) -- (3.4,0.7);
\draw[black, line width=1] (3.4,0) -- (3.4,0.7);

\draw[black, line width=1] (5.4,0) -- (5.4,0.7);
\draw[black, line width=1] (5.4,0) -- (6.1,0);
\draw[black, line width=1] (5.4,0.7) -- (6.1,0.7);
\draw[black, line width=1] (6.1,0) -- (6.1,0.7);

\draw[black, line width=1] (2.7,2.7) -- (2.7,3.4);
\draw[black, line width=1] (2.7,2.7) -- (3.4,2.7);
\draw[black, line width=1] (2.7,3.4) -- (3.4,3.4);
\draw[black, line width=1] (3.4,2.7) -- (3.4,3.4);

\draw[black, line width=1] (5.4,2.7) -- (5.4,3.4);
\draw[black, line width=1] (5.4,2.7) -- (6.1,2.7);
\draw[black, line width=1] (5.4,3.4) -- (6.1,3.4);
\draw[black, line width=1] (6.1,2.7) -- (6.1,3.4);

\draw[black, line width=1] (0,2.7) -- (0,3.4);
\draw[black, line width=1] (0,2.7) -- (0.7,2.7);
\draw[black, line width=1] (0,3.4) -- (0.7,3.4);
\draw[black, line width=1] (0.7,2.7) -- (0.7,3.4);


\draw[black, line width=1] (0.7,0.35) -- (2.7,0.35);
\draw[black, line width=1] (5.75,0.7) -- (5.75,2.7);
\draw[black, line width=1] (0.7,3.05) -- (2.7,3.05);
\draw[black, line width = 1] (3.05,0.7) -- (3.05,2.7);
\draw[black, line width = 1] (3.05,0.7) -- (0.35,2.7);

\draw[black, gluon, line width=1] (3.4,0.35) -- (5.4,0.35);
\draw[black, gluon, line width=1] (3.4,3.05) -- (5.4,3.05);

\draw[red, dashed, line width = 1] (0.35,0.7) -- (0.35,2.7);

\draw[violet,line width = 1,->] (0.35,1.5) -- (1.5,1);
\filldraw (1.7,1) node {$\emptyset$};
\filldraw (1,1.5) node {$t$};

\draw[violet,line width = 1,->] (1.3,3.05) -- (1.7,2.6);
\filldraw (1.85,2.35) node {$\{1\}$};
\draw[violet,line width = 1,->] (3.05,1.6) -- (2.2,2.2);
\filldraw (1.3,2.7) node {$t$};
\filldraw (2.7,2.1) node {$t$};

\draw[violet,line width = 1,->] (3.7,3.05) -- (4.3,2.2);
\filldraw (4.3,1.9) node {$\{2\}$};
\draw[violet,line width = 1,->] (3.7,0.35) -- (4.3,1.6);
\filldraw (3.8,2.6) node {$t$};
\filldraw (4.2,1) node {$t$};





\end{tikzpicture}
}
\hspace{0.1in}
\subfigure[]{
\begin{tikzpicture}
\filldraw (0.35,0.35) node {$V^{[1]}_4$};
\filldraw (5.75,3.05) node {$V^{[1]}_3$};
\filldraw (3.05,3.05) node {$V^{[1]}_2$};
\filldraw (0.35,3.05) node {$V^{[1]}_1$};
\filldraw (5.75,0.35) node {$V^{[1]}_6$};
\filldraw (3.05,0.35) node {$V^{[1]}_5$};

\filldraw (0.35,3.75) node {$\textcolor{red}{W_1}+Z$};
\filldraw (0.35,-0.35) node {$\textcolor{red}{W_1}+Z$};

\filldraw (5.75,3.75) node {$\textcolor{blue}{2W_1}+Z$};
\filldraw (3.05,3.75) node {$\textcolor{blue}{2W_1}+Z$};

\filldraw (5.75,-0.35) node {$\textcolor{violet}{3W_1}+Z$};
\filldraw (3.05,-0.35) node {$\textcolor{violet}{3W_1}+Z$};

\draw[black, line width=1] (0,0) -- (0,0.7);
\draw[black, line width=1] (0,0) -- (0.7,0);
\draw[black, line width=1] (0,0.7) -- (0.7,0.7);
\draw[black, line width=1] (0.7,0) -- (0.7,0.7);

\draw[black, line width=1] (2.7,0) -- (2.7,0.7);
\draw[black, line width=1] (2.7,0) -- (3.4,0);
\draw[black, line width=1] (2.7,0.7) -- (3.4,0.7);
\draw[black, line width=1] (3.4,0) -- (3.4,0.7);

\draw[black, line width=1] (5.4,0) -- (5.4,0.7);
\draw[black, line width=1] (5.4,0) -- (6.1,0);
\draw[black, line width=1] (5.4,0.7) -- (6.1,0.7);
\draw[black, line width=1] (6.1,0) -- (6.1,0.7);

\draw[black, line width=1] (2.7,2.7) -- (2.7,3.4);
\draw[black, line width=1] (2.7,2.7) -- (3.4,2.7);
\draw[black, line width=1] (2.7,3.4) -- (3.4,3.4);
\draw[black, line width=1] (3.4,2.7) -- (3.4,3.4);

\draw[black, line width=1] (5.4,2.7) -- (5.4,3.4);
\draw[black, line width=1] (5.4,2.7) -- (6.1,2.7);
\draw[black, line width=1] (5.4,3.4) -- (6.1,3.4);
\draw[black, line width=1] (6.1,2.7) -- (6.1,3.4);

\draw[black, line width=1] (0,2.7) -- (0,3.4);
\draw[black, line width=1] (0,2.7) -- (0.7,2.7);
\draw[black, line width=1] (0,3.4) -- (0.7,3.4);
\draw[black, line width=1] (0.7,2.7) -- (0.7,3.4);


\draw[black, line width=1] (0.7,0.35) -- (2.7,0.35);
\draw[black, line width=1] (5.75,0.7) -- (5.75,2.7);
\draw[black, line width=1] (0.7,3.05) -- (2.7,3.05);
\draw[black, line width = 1] (3.05,0.7) -- (3.05,2.7);
\draw[black, line width = 1] (3.05,0.7) -- (0.35,2.7);

\draw[red, dashed, line width=1] (3.4,0.35) -- (5.4,0.35);
\draw[red, dashed, line width=1] (3.4,3.05) -- (5.4,3.05);

\draw[red, dashed, line width = 1] (0.35,0.7) -- (0.35,2.7);

\draw[violet,line width = 1,->] (1.3,3.05) -- (1.7,2.6);
\filldraw (1.85,2.35) node {$\{1\}$};
\draw[violet,line width = 1,->] (3.05,1.6) -- (2.2,2.2);
\filldraw (1.3,2.7) node {$t$};
\filldraw (2.7,2.1) node {$t$};

\draw[violet,line width = 1,->] (3.7,3.05) -- (4.3,2.2);
\filldraw (4.3,1.9) node {$\emptyset$};
\draw[violet,line width = 1,->] (3.7,0.35) -- (4.3,1.6);
\filldraw (3.8,2.6) node {$t$};
\filldraw (4.2,1) node {$t$};



\end{tikzpicture}
}

\end{center}
\vspace{-0.25in}
  \caption{\small (a) An example graph $G$ and (b) its $G^{[1]}$ of $W_1$. The secure storage capacity over $G$ is $1$ and a capacity achieving code construction is shown.
  }
  \label{fig:ex1}
\end{figure}

\begin{example}\label{ex1}
Consider the secure storage problem instance in Fig.~\ref{fig:ex1}. Each node has security constraint such that $L_v \geq L_w$ and $R \leq 1$. An optimal code with $R=1$ is constructed as follows. Suppose each coded symbol $W_k$ is from $\mathbb{F}_5$. First, $G$ is a qualified component so that the same independent noise variable $Z$ (uniform over $\mathbb{F}_5$) must be used (called noise alignment, refer to Lemma \ref{lemma:noise}). Second, the coded symbols are designed by considering each $W_k$ and $G^{[k]}$ separately. For example, consider $W_1$ and $G^{[1]}$ in Fig.~\ref{fig:ex1}.(b), wherein an unqualified component cannot reveal anything about $W_1$ so that the same coded symbol must be assigned (called coded symbol alignment, refer to Lemma \ref{lemma:signal}). We then assign a generic combination to each unqualified component, e.g., $V_1^{[1]} = V_4^{[1]} = W_1+Z, V_2^{[1]} = V_3^{[1]} = 2W_1+Z, V_5^{[1]} = V_6^{[1]} = 3W_1+Z$ (colored differently in Fig.~\ref{fig:ex1}.(b)). As there is no internal qualified edge, all qualified edges span different unqualified components and contain linearly independent combinations of the source symbol and the noise, from which the desired source symbol can be obtained (e.g., see $(V_1^{[1]}, V_5^{[1]})$ in Fig.~\ref{fig:ex1}.(b)). Finally, we combine (add) the source symbol assignment in each $G^{[k]}$ to produce the coded symbol assignment in $G$ so that for any edge, the desired source symbol has different coefficients (thus correct) and undesired source symbols (and noise) are aligned (thus secure).
\end{example}

The second example (see Fig.~\ref{fig:ex2}) is used to explain the `only if' part, i.e., the graph $G$ does not satisfy the condition in Theorem \ref{thm:d1} 
and the secure graph capacity $C < 1$.

 \tikzset{
    photon/.style={decorate, decoration={snake,
      amplitude = 0.5mm,
      segment length = 3mm}, draw=green},
    electron/.style={draw=blue, postaction={decorate},
        decoration={markings,mark=at position .55 with {\arrow[draw=blue]{>}}}},
    gluon/.style={decorate, draw=blue,
        decoration={coil,amplitude=1.5pt, segment length=4pt}} 
}
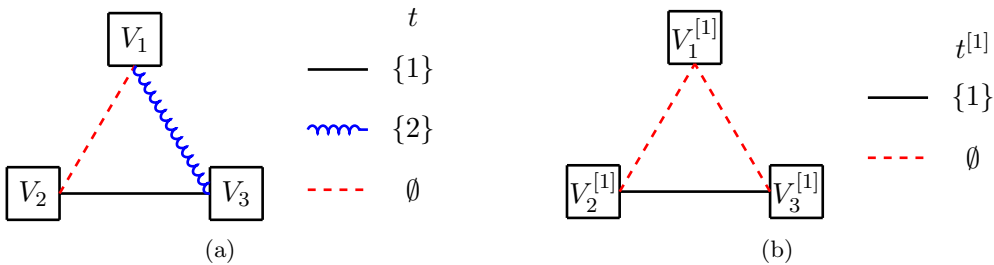
\begin{figure}[H]
\centering
\subfigure[]{
\begin{tikzpicture}
\filldraw (1.7,2.4) node {$V_1$};
\filldraw (0.35,0.35) node {$V_2$};
\filldraw (3.05,0.35) node {$V_3$};

\draw[black, line width=1] (0,0) -- (0,0.7);
\draw[black, line width=1] (0,0) -- (0.7,0);
\draw[black, line width=1] (0,0.7) -- (0.7,0.7);
\draw[black, line width=1] (0.7,0) -- (0.7,0.7);

\draw[black, line width=1] (2.7,0) -- (2.7,0.7);
\draw[black, line width=1] (2.7,0) -- (3.4,0);
\draw[black, line width=1] (2.7,0.7) -- (3.4,0.7);
\draw[black, line width=1] (3.4,0) -- (3.4,0.7);

\draw[black, line width=1] (1.35,2.05) -- (1.35,2.75);
\draw[black, line width=1] (1.35,2.05) -- (2.05,2.05);
\draw[black, line width=1] (1.35,2.75) -- (2.05,2.75);
\draw[black, line width=1] (2.05,2.05) -- (2.05,2.75);

\draw[black, line width=1] (0.7,0.35) -- (2.7,0.35);
\draw[red, dashed, line width = 1](0.7,0.35) -- (1.7,2.05);
\draw[black, gluon, line width=1](2.7,0.35) -- (1.7,2.05);

\draw[black, line width=1] (4,2) -- (4.8,2);
\draw[black, gluon, line width=1](4,1.2) -- (4.8,1.2);
\draw[red, dashed, line width=1](4,0.4) -- (4.8,0.4);

\filldraw (5.4, 2.7) node {$t$};
\filldraw (5.4, 2) node {$\{1\}$};
\filldraw (5.4, 1.2) node {$\{2\}$};
\filldraw (5.4, 0.4) node {$\emptyset$};

\end{tikzpicture}
}
\hspace{0.4in}
\subfigure[]{
\begin{tikzpicture}
\filldraw (1.7,2.4) node {$V^{[1]}_1$};
\filldraw (0.35,0.35) node {$V^{[1]}_2$};
\filldraw (3.05,0.35) node {$V^{[1]}_3$};

\draw[black, line width=1] (0,0) -- (0,0.7);
\draw[black, line width=1] (0,0) -- (0.7,0);
\draw[black, line width=1] (0,0.7) -- (0.7,0.7);
\draw[black, line width=1] (0.7,0) -- (0.7,0.7);

\draw[black, line width=1] (2.7,0) -- (2.7,0.7);
\draw[black, line width=1] (2.7,0) -- (3.4,0);
\draw[black, line width=1] (2.7,0.7) -- (3.4,0.7);
\draw[black, line width=1] (3.4,0) -- (3.4,0.7);

\draw[black, line width=1] (1.35,2.05) -- (1.35,2.75);
\draw[black, line width=1] (1.35,2.05) -- (2.05,2.05);
\draw[black, line width=1] (1.35,2.75) -- (2.05,2.75);
\draw[black, line width=1] (2.05,2.05) -- (2.05,2.75);

\draw[black, line width=1] (0.7,0.35) -- (2.7,0.35);
\draw[red, dashed, line width = 1](0.7,0.35) -- (1.7,2.05);
\draw[red, dashed,line width=1](2.7,0.35) -- (1.7,2.05);


\draw[black, line width=1] (4,1.6) -- (4.8,1.6);
\draw[red, dashed, line width=1](4,0.8) -- (4.8,0.8);

\filldraw (5.4, 2.3) node {$t^{[1]}$};
\filldraw (5.4, 1.6) node {$\{1\}$};
\filldraw (5.4, 0.8) node {$\emptyset$};

\end{tikzpicture}
}
\vspace{-0.1in}
  \caption{\small (a) An example graph $G$ and (b) its $G^{[1]}$ of $W_1$. The secure storage capacity over $G$ cannot be $1$.}
\label{fig:ex2}
\end{figure}

\begin{example}\label{ex2}
Consider the secure storage problem instance in Fig.~\ref{fig:ex2}. $G^{[1]}$ contains an internal qualified edge $\{V_2^{[1]}, V_3^{[1]}\}$ inside the unqualified path $\left(\{V_2^{[1]}, V_1^{[1]}\}, \{V_1^{[1]}, V_3^{[1]}\}\right)$. The intuition that $C \neq 1$, i.e., $L_v \neq L_w$ is as follows (ignoring $o(L_w)$ terms). When $L_v = L_w$, in Fig.~\ref{fig:ex2}.(a), $G$ is a qualified component so that the same noise must be used in each of $V_1, V_2, V_3$ (called noise alignment, refer to Lemma \ref{lemma:noise}); in Fig.~\ref{fig:ex2}.(b), $\left(\{V_2^{[1]}, V_1^{[1]}\}, \{V_1^{[1]}, V_3^{[1]}\}\right)$ is an unqualified path so that the same coded symbol about $W_1$ must be stored in $V_1, V_2, V_3$ (called coded symbol alignment, which can be captured by conditioned entropy. Refer to Lemma \ref{lemma:signal}). It then follows that $V_2, V_3$ must store the same information about $W_1$, which contradicts the fact that from $(V_2, V_3)$, we can decode $W_1$. The above discussion can be translated to entropy manipulations and the details are presented in Section \ref{sec:thm21}.
\end{example}

\subsection{Arbitrary $D$ and Extremal Graphs with $C=1/D$}
Next, we extend Theorem \ref{thm:d1} to the setting of arbitrary $D$. Under the condition that every non-degenerate node has no common source, all extremal graphs whose secure storage capacity is $C=1/D$ are characterized in the following theorem.

\begin{theorem}\label{thm:d2}
Consider the class of graph $G = (\mathcal{V}, \mathcal{E})$ where the non-degenerate subgraph $\widetilde{G}$ of $G$ is not empty and $\mathcal{C}(V)=\emptyset, \forall V\in\mathcal{V}\backslash\mathcal{V}_d$.
For this class of graph $G$, the capacity of a secure storage code 
is $C=1/D$ if and only if for every qualified component $Q$ of $\widetilde{G}$, the characteristic graph $Q^{[k]}$ of each coded symbol $W_k, k \in [K]$ contains no internal qualified edge.
\end{theorem}

\begin{remark}
Theorem \ref{thm:d2} includes Theorem \ref{thm:d1} as a special case because when $D=1$, any non-degenerate node must have no common source.
\end{remark}

The proof of Theorem \ref{thm:d2} is presented in Section \ref{sec:thm2}. An example is given  in Fig.~\ref{fig:ex3} to explain the code construction of the `if' part. 

\tikzset{
    photon/.style={decorate, decoration={snake,
      amplitude = 0.5mm,
      segment length = 3mm}, draw=violet},
    electron/.style={draw=blue, postaction={decorate},
        decoration={markings,mark=at position .55 with {\arrow[draw=blue]{>}}}},
    gluon/.style={decorate, draw=blue,
        decoration={coil,amplitude=1.5pt, segment length=4pt}} 
}
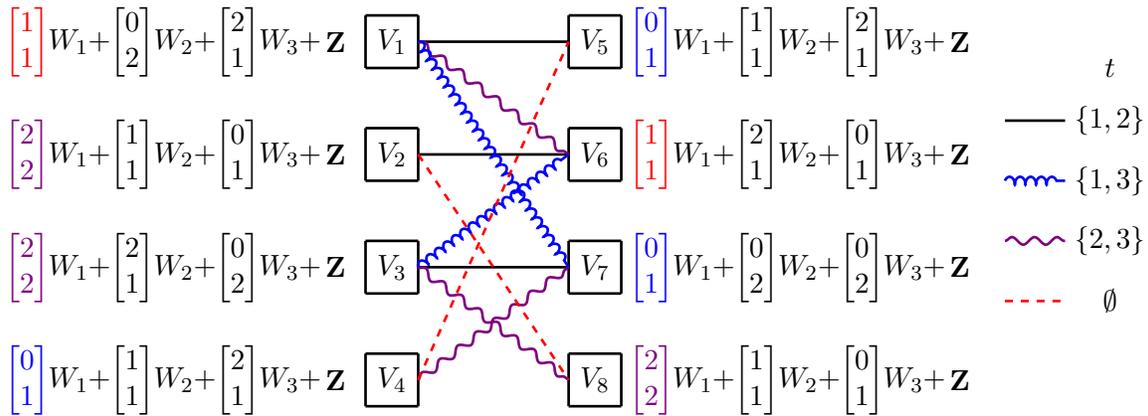
\begin{figure}[H]
\begin{center}
\begin{tikzpicture}
\filldraw (1.35,-1.15) node {$V_4$};
\filldraw (1.35,0.35) node {$V_3$};
\filldraw (1.35,1.85) node {$V_2$};
\filldraw (1.35,3.35) node {$V_1$};
\filldraw (4.05,-1.15) node {$V_8$};
\filldraw (4.05,0.35) node {$V_7$};
\filldraw (4.05,1.85) node {$V_6$};
\filldraw (4.05,3.35) node {$V_5$};

\node (def) at (-3.5,3.35){$\textcolor{red}{\begin{bmatrix}{}
1\\
1
\end{bmatrix}}
$};
\filldraw (-2.8,3.35) node {$W_1+$};

\node (def) at (-2.1,3.35){$\begin{bmatrix}{}
0\\
2
\end{bmatrix}
$};
\filldraw (-1.4,3.35) node {$W_2+$};

\node (def) at (-0.7,3.35){$\begin{bmatrix}{}
2\\
1
\end{bmatrix}
$};
\filldraw (0,3.35) node {$W_3+$};

\filldraw (0.6,3.35) node {${\bf Z}$};

\node (def) at (4.8,3.35){$\textcolor{blue}{\begin{bmatrix}{}
0\\
1
\end{bmatrix}}
$};
\filldraw (5.5,3.35) node {$W_1+$};

\node (def) at (6.2,3.35){$\begin{bmatrix}{}
1\\
1
\end{bmatrix}
$};
\filldraw (6.9,3.35) node {$W_2+$};

\node (def) at (7.6,3.35){$\begin{bmatrix}{}
2\\
1
\end{bmatrix}
$};
\filldraw (8.3,3.35) node {$W_3+$};

\filldraw (8.9,3.35) node {${\bf Z}$};

\node (def) at (-3.5,1.85){$\textcolor{violet}{\begin{bmatrix}{}
2\\
2
\end{bmatrix}}
$};
\filldraw (-2.8,1.85) node {$W_1+$};

\node (def) at (-2.1,1.85){$\begin{bmatrix}{}
1\\
1
\end{bmatrix}
$};
\filldraw (-1.4,1.85) node {$W_2+$};

\node (def) at (-0.7,1.85){$\begin{bmatrix}{}
0\\
1
\end{bmatrix}
$};
\filldraw (0,1.85) node {$W_3+$};

\filldraw (0.6,1.85) node {${\bf Z}$};

\node (def) at (4.8,1.85){$\textcolor{red}{\begin{bmatrix}{}
1\\
1
\end{bmatrix}}
$};
\filldraw (5.5,1.85) node {$W_1+$};

\node (def) at (6.2,1.85){$\begin{bmatrix}{}
2\\
1
\end{bmatrix}
$};
\filldraw (6.9,1.85) node {$W_2+$};

\node (def) at (7.6,1.85){$\begin{bmatrix}{}
0\\
1
\end{bmatrix}
$};
\filldraw (8.3,1.85) node {$W_3+$};

\filldraw (8.9,1.85) node {${\bf Z}$};

\node (def) at (-3.5,0.35){$\textcolor{violet}{\begin{bmatrix}{}
2\\
2
\end{bmatrix}}
$};
\filldraw (-2.8,0.35) node {$W_1+$};

\node (def) at (-2.1,0.35){$\begin{bmatrix}{}
2\\
1
\end{bmatrix}
$};
\filldraw (-1.4,0.35) node {$W_2+$};

\node (def) at (-0.7,0.35){$\begin{bmatrix}{}
0\\
2
\end{bmatrix}
$};
\filldraw (0,0.35) node {$W_3+$};

\filldraw (0.6,0.35) node {${\bf Z}$};

\node (def) at (4.8,0.35){$\textcolor{blue}{\begin{bmatrix}{}
0\\
1
\end{bmatrix}}
$};
\filldraw (5.5,0.35) node {$W_1+$};

\node (def) at (6.2,0.35){$\begin{bmatrix}{}
0\\
2
\end{bmatrix}
$};
\filldraw (6.9,0.35) node {$W_2+$};

\node (def) at (7.6,0.35){$\begin{bmatrix}{}
0\\
2
\end{bmatrix}
$};
\filldraw (8.3,0.35) node {$W_3+$};

\filldraw (8.9,0.35) node {${\bf Z}$};

\node (def) at (-3.5,-1.15){$\textcolor{blue}{\begin{bmatrix}{}
0\\
1
\end{bmatrix}}
$};
\filldraw (-2.8,-1.15) node {$W_1+$};

\node (def) at (-2.1,-1.15){$\begin{bmatrix}{}
1\\
1
\end{bmatrix}
$};
\filldraw (-1.4,-1.15) node {$W_2+$};

\node (def) at (-0.7,-1.15){$\begin{bmatrix}{}
2\\
1
\end{bmatrix}
$};
\filldraw (0,-1.15) node {$W_3+$};

\filldraw (0.6,-1.15) node {${\bf Z}$};

\node (def) at (4.8,-1.15){$\textcolor{violet}{\begin{bmatrix}{}
2\\
2
\end{bmatrix}}
$};
\filldraw (5.5,-1.15) node {$W_1+$};

\node (def) at (6.2,-1.15){$\begin{bmatrix}{}
1\\
1
\end{bmatrix}
$};
\filldraw (6.9,-1.15) node {$W_2+$};

\node (def) at (7.6,-1.15){$\begin{bmatrix}{}
0\\
1
\end{bmatrix}
$};
\filldraw (8.3,-1.15) node {$W_3+$};

\filldraw (8.9,-1.15) node {${\bf Z}$};

\draw[black, line width=1] (1,-1.5) -- (1,-0.8);
\draw[black, line width=1] (1,-1.5) -- (1.7,-1.5);
\draw[black, line width=1] (1,-0.8) -- (1.7,-0.8);
\draw[black, line width=1] (1.7,-1.5) -- (1.7,-0.8);

\draw[black, line width=1] (3.7,-1.5) -- (3.7,-0.8);
\draw[black, line width=1] (3.7,-1.5) -- (4.4,-1.5);
\draw[black, line width=1] (3.7,-0.8) -- (4.4,-0.8);
\draw[black, line width=1] (4.4,-1.5) -- (4.4,-0.8);

\draw[black, line width=1] (1,0) -- (1,0.7);
\draw[black, line width=1] (1,0) -- (1.7,0);
\draw[black, line width=1] (1,0.7) -- (1.7,0.7);
\draw[black, line width=1] (1.7,0) -- (1.7,0.7);

\draw[black, line width=1] (3.7,0) -- (3.7,0.7);
\draw[black, line width=1] (3.7,0) -- (4.4,0);
\draw[black, line width=1] (3.7,0.7) -- (4.4,0.7);
\draw[black, line width=1] (4.4,0) -- (4.4,0.7);

\draw[black, line width=1] (1,1.5) -- (1,2.2);
\draw[black, line width=1] (1,1.5) -- (1.7,1.5);
\draw[black, line width=1] (1,2.2) -- (1.7,2.2);
\draw[black, line width=1] (1.7,1.5) -- (1.7,2.2);

\draw[black, line width=1] (3.7,1.5) -- (3.7,2.2);
\draw[black, line width=1] (3.7,1.5) -- (4.4,1.5);
\draw[black, line width=1] (3.7,2.2) -- (4.4,2.2);
\draw[black, line width=1] (4.4,1.5) -- (4.4,2.2);

\draw[black, line width=1] (1,3) -- (1,3.7);
\draw[black, line width=1] (1,3) -- (1.7,3);
\draw[black, line width=1] (1,3.7) -- (1.7,3.7);
\draw[black, line width=1] (1.7,3) -- (1.7,3.7);

\draw[black, line width=1] (3.7,3) -- (3.7,3.7);
\draw[black, line width=1] (3.7,3) -- (4.4,3);
\draw[black, line width=1] (3.7,3.7) -- (4.4,3.7);
\draw[black, line width=1] (4.4,3) -- (4.4,3.7);

\draw[black, line width=1] (1.7,3.35) -- (3.7,3.35);
\draw[black, line width=1] (1.7,1.85) -- (3.7,1.85);
\draw[black, line width=1] (1.7,0.35) -- (3.7,0.35);

\draw[green, photon, line width=1] (1.7,3.35) -- (3.7,1.85);
\draw[green, photon, line width=1] (1.7,0.35) -- (3.7,-1.15);
\draw[green, photon, line width=1] (1.7,-1.15) -- (3.7,0.35);

\draw[black, gluon, line width=1] (1.7,3.35) -- (3.7,0.35);
\draw[black, gluon, line width=1] (1.7,0.35) -- (3.7,1.85);

\draw[red, dashed, line width = 1] (1.7,1.85) -- (3.7,-1.15);
\draw[red, dashed, line width = 1] (1.7,-1.15) -- (3.7,3.35);

\draw[black, line width=1] (9.5, 2.3) -- (10.3,2.3);
\draw[black, gluon, line width=1](9.5, 1.5) -- (10.3, 1.5);
\draw[black, photon, line width=1](9.5, 0.7) -- (10.3, 0.7);
\draw[red, dashed, line width=1](9.5, -0.1) -- (10.3,-0.1);

\filldraw (10.9,3) node {$t$};
\filldraw (10.9,2.3) node {$\{1,2\}$};
\filldraw (10.9,1.5) node {$\{1,3\}$};
\filldraw (10.9,0.7) node {$\{2,3\}$};
\filldraw (10.9,-0.1) node {$\emptyset$};

\end{tikzpicture}
\end{center}
\vspace{-0.2in}
  \caption{\small An example graph $G$ and a code construction that achieves the capacity $1/2$.
  }
  \label{fig:ex3}
\end{figure}

\begin{example}
Consider the graph $G$ in Fig.~\ref{fig:ex3}. The code construction is based on a similar idea as that of Example \ref{ex1}, i.e., each $W_k$ and $G^{[k]}$ is considered separately and generic combinations are assigned to each unqualified component of $G^{[k]}$ (colored differently in Fig.~\ref{fig:ex3} for $W_1$); then the overall assignment is obtained as the sum of each assignment in $G^{[k]}$. In Fig.~\ref{fig:ex3}, each $W_k$ is from $\mathbb{F}_3$ and ${\bf Z} \in \mathbb{F}_3^{2\times 1}$ is an independent uniform noise. The main difference between this example where $D=2$ and Example \ref{ex1} where $D=1$ is that to ensure correctness, $D=1$ only requires the coefficients of the desired source symbol to be different while $D>1$ needs the coefficient matrix to be full rank (for which an explicit design is not obvious). An explicit solution is provided in Fig.~\ref{fig:ex3} for this small example while in general, the proof in Section \ref{sec:thm22} relies on randomized construction. 
\end{example}

\subsection{Arbitrary $D$ and Extremal Graphs with $C=2/D$}
Finally, we consider the extremal rate of $2/D$. All extremal graphs whose secure storage capacity is $C=2/D$ are characterized in the following theorem.

\begin{theorem}\label{thm:2d}
The capacity of a secure storage code over a graph $G = (\mathcal{V}, \mathcal{E})$ is $C=2/D$ if and only the following two conditions are satisfied.
\begin{enumerate}
    \item For any $V \in \mathcal{V}$, $|\mathcal{C}(V)| \geq D/2$. 
    \item For any $\{V_i, V_j\} \in \mathcal{E}$, 
    $\mathcal{C}(V_i) \cup \mathcal{C}(V_j) = t(\{V_i, V_j\})$.
\end{enumerate}
\end{theorem}

In words, the conditions in Theorem \ref{thm:2d} are 1). for each node, there are at least $D/2$ common sources and 2). for each qualified edge, the union of the common sources of both nodes must be the set of $D$ desired source symbols. The intuition is fairly straightforward as the total storage of any qualified edge is exactly $2L_v = 2 \times 1/R \times L_w = D \times L_w$ {\color{black}(ignoring $o(L_w)$ terms)}, which must be fully occupied by the desired $D$ source symbols and there is absolutely no room for anything else. As a consequence, we can show that each coded symbol must be a deterministic function of its common sources (refer to Lemma \ref{lemma:det}). Then the two conditions in Theorem \ref{thm:2d} follow as necessary conditions as otherwise we do not have sufficient information from the desired source symbols to fill a node and a qualified edge. 
The two conditions also turn out to be sufficient by random linear coding, i.e., storing a sufficient number of generic combinations of the common sources guarantees the successful recovery of the desired source symbols (see Fig.~\ref{fig:ex4} for an example). The detailed proof of Theorem \ref{thm:2d} is deferred to Section \ref{sec:thm3}.

\tikzset{
    photon/.style={decorate, decoration={snake,
      amplitude = 0.5mm,
      segment length = 3mm}, draw=violet},
    electron/.style={draw=blue, postaction={decorate},
        decoration={markings,mark=at position .55 with {\arrow[draw=blue]{>}}}},
    gluon/.style={decorate, draw=blue,
        decoration={coil,amplitude=1.5pt, segment length=4pt}} 
}
\begin{figure}[H]
\begin{center}
\begin{tikzpicture}
\filldraw (0.35,0.35) node {$V_4$};
\filldraw (5.05,2.65) node {$V_3$};
\filldraw (2.65,2.65) node {$V_2$};
\filldraw (0.35,2.65) node {$V_1$};
\filldraw (5.05,0.35) node {$V_6$};
\filldraw (2.65,0.35) node {$V_5$};

\filldraw (0.35,3.35) node {$W_1$};
\filldraw (2.65,3.35) node {$W_1+W_3$};
\filldraw (5.05,3.35) node {$W_2$};


\filldraw (0.35,-0.35) node {$W_1+W_2$};
\filldraw (2.65,-0.35) node {$W_1+2W_2$};
\filldraw (5.05,-0.35) node {$W_3$};

\draw[black, line width=1] (0,0) -- (0,0.7);
\draw[black, line width=1] (0,0) -- (0.7,0);
\draw[black, line width=1] (0,0.7) -- (0.7,0.7);
\draw[black, line width=1] (0.7,0) -- (0.7,0.7);

\draw[black, line width=1] (2.3,0) -- (2.3,0.7);
\draw[black, line width=1] (2.3,0) -- (3,0);
\draw[black, line width=1] (2.3,0.7) -- (3,0.7);
\draw[black, line width=1] (3,0) -- (3,0.7);

\draw[black, line width=1] (4.7,0) -- (4.7,0.7);
\draw[black, line width=1] (4.7,0) -- (5.4,0);
\draw[black, line width=1] (4.7,0.7) -- (5.4,0.7);
\draw[black, line width=1] (5.4,0) -- (5.4,0.7);

\draw[black, line width=1] (2.3,2.3) -- (2.3,3);
\draw[black, line width=1] (2.3,2.3) -- (3,2.3);
\draw[black, line width=1] (2.3,3) -- (3,3);
\draw[black, line width=1] (3,2.3) -- (3,3);

\draw[black, line width=1] (4.7,2.3) -- (4.7,3);
\draw[black, line width=1] (4.7,2.3) -- (5.4,2.3);
\draw[black, line width=1] (4.7,3) -- (5.4,3);
\draw[black, line width=1] (5.4,2.3) -- (5.4,3);

\draw[black, line width=1] (0,2.3) -- (0,3);
\draw[black, line width=1] (0,2.3) -- (0.7,2.3);
\draw[black, line width=1] (0,3) -- (0.7,3);
\draw[black, line width=1] (0.7,2.3) -- (0.7,3);


\draw[black, line width = 1] (0.35,0.7) -- (0.35,2.3);
\draw[black, line width=1] (0.7,0.35) -- (2.3,0.35);
\draw[black, line width = 1] (2.65,0.7) -- (5.05,2.3);
\draw[black, line width = 1] (2.65,0.7) -- (0.35,2.3);
\draw[black, line width = 1] (0.35,0.7) -- (5.05,2.3);

\draw[black, gluon, line width=1] (0.7,2.65) -- (2.3,2.65);
\draw[black, gluon, line width=1] (2.65,2.3) -- (5.05,0.7);

\draw[black, photon, line width=1](5.05,0.7) -- (5.05,2.3);

\draw[black, line width=1] (6.2,2.35) -- (7,2.35);
\draw[black, gluon, line width=1](6.2,1.55) -- (7,1.55);
\draw[black, photon, line width=1](6.2,0.75) -- (7,0.75);


\filldraw (7.7, 2.85) node {$t$};
\filldraw (7.7, 2.35) node {$\{1,2\}$};
\filldraw (7.7, 1.55) node {$\{1,3\}$};
\filldraw (7.7, 0.75) node {$\{2,3\}$};

\end{tikzpicture}
\end{center}
\vspace{-0.2in}
  \caption{\small An example graph $G$ and a code construction that achieves the capacity $1$.
  }
  \label{fig:ex4}
\end{figure}
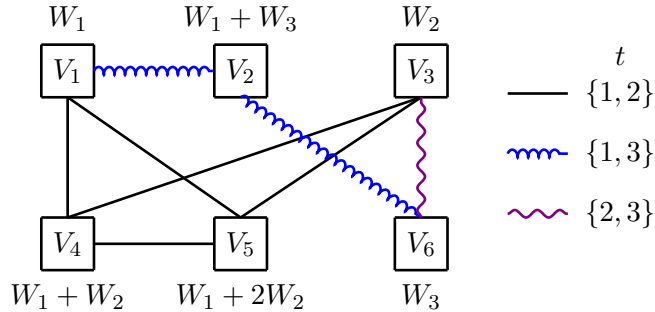

\section{Proof of Theorem \ref{thm:d1}}\label{sec:thm1}
Theorem \ref{thm:d1} is recovered as a special case of Theorem \ref{thm:d2}, so the proof of Theorem \ref{thm:d1} can also be recovered from the proof of Theorem \ref{thm:d2}, presented in Section \ref{sec:thm2}. However, in this section we still provide a proof of the code construction for the `if' part because here $D=1$, the code can be made explicit while in Theorem \ref{thm:d2} where $D$ can be arbitrary, the code is randomized.

\subsection{Code Construction of the `if' Part}\label{sec:thm11}
We show that if $G$ satisfies the condition in Theorem \ref{thm:d1}, then we can construct a secure storage code of symbol rate $R = 1$. Suppose $L_w = \log_2(q)$ bits and each source symbol $W_k$ is one symbol from finite field $\mathbb{F}_q$, where the field size $q$ will be specified later in the proof. Each coded symbol $V_n$ will be set as one symbol from $\mathbb{F}_q$, i.e., $L_v = \log_2(q)$ bits and $R = L_w/L_v = 1$, as desired.

A degenerate node $V\in\mathcal{V}_d$, i.e., all connected edges are associated with the same coded symbol $W_i$ (or all connected edges are unqualified, in this case set $V$ to contain an independent noise variable), is trivial, and set $V = W_i$. Henceforth, we focus on the non-degenerate subgraph $\widetilde{G}$ of $G$, i.e., all non-degenerate nodes $\mathcal{V}\backslash\mathcal{V}_d$. Suppose $\widetilde{G}$ has $M$ qualified components, $Q_1, \cdots, Q_M$. 
To assign the coded symbols in $Q_m, m\in [M]$, we will first consider the characteristic graph $Q_m^{[k]}, m \in [M], k \in [K]$ of each coded symbol $W_k$ separately and then combine the separated assignments. 

{\bf Consider $Q_m^{[k]}, \forall m \in [M], k \in [K]$.} Suppose $Q_m^{[k]}$ contains $U_m^{[k]}$ unqualified components and set the field size $q$ as a prime number such that $q > \max_{m,k} U_m^{[k]}$. The nodes in $Q_m^{[k]}$ are set as follows.
\begin{eqnarray}
&& \mbox{For each node $V^{[k]}$ in the $u$-th unqualified component of $Q_m^{[k]}$ where $u\in[U_m^{[k]}]$,}\notag\\
&& ~~\mbox{set $V^{[k]} = u \times W_k +  Z_m^{[k]}$} \label{eq:c1}
\end{eqnarray}
where $Z_m^{[k]}, \forall m, k$ are i.i.d. uniform noise symbols from $\mathbb{F}_q$ and are independent of $W_k$. As the condition of Theorem \ref{thm:d1} is satisfied, i.e., $Q_m^{[k]}$ contains no internal qualified edge, the assignment (\ref{eq:c1}) satisfies the following two properties.
\begin{eqnarray}
&& \mbox{For each qualified edge $\{V_i^{[k]}, V_j^{[k]}\}$ in $Q_m^{[k]}$, from $V_i^{[k]} - V_j^{[k]}$ we can obtain $W_k$.} \label{eq:p1}\\
&& \mbox{For each unqualified edge $\{V_i^{[k]}, V_j^{[k]}\}$ in $Q_m^{[k]}$, $V_i^{[k]} = V_j^{[k]}$.} \label{eq:p2}
\end{eqnarray}
(\ref{eq:p1}) follows from the observation that as there is no internal edge, $V_i^{[k]}$ and $V_j^{[k]}$ belong to different unqualified components such that the coefficients before $W_k$ are different (see (\ref{eq:c1})). (\ref{eq:p2}) follows from the fact that any unqualified edge belongs to the same unqualified component and (\ref{eq:c1}).

{\bf Consider $Q_m, \forall m \in [M]$.} Note that the nodes in $Q_m$ and $Q_m^{[k]}$ have a one-to-one mapping. Each node in $Q_m$ is simply set as the sum of each corresponding node in $Q_m^{[k]}$ for all $k \in [K]$.
\begin{eqnarray}
\mbox{For each node $V$ in $Q_m$, set $V = \sum_{k\in[K]} V^{[k]}$.} \label{eq:c2}
\end{eqnarray}

{\bf We show that the code construction (\ref{eq:c1}), (\ref{eq:c2}) is correct and secure.} Edges connected to degenerate nodes are trivial and we only need to consider the remaining edges. Consider any qualified edge $\{V_i, V_j\}$, i.e., $t(\{V_i,V_j\}) = \{l\}$.
\begin{eqnarray}
V_i - V_j &\overset{(\ref{eq:c2})}{=}& \sum_{k\in[K]} \left( V^{[k]}_i - V^{[k]}_j \right) \\
&\overset{(\ref{eq:p2})}{=}& V^{[l]}_i - V^{[l]}_j
\end{eqnarray}
where the last step follows from the fact that $\{V^{[k]}_i, V^{[k]}_j\}, k \neq l$ is an unqualified edge as $t(\{V_i,V_j\}) = \{l\}$ 
(see Definition \ref{def:cha}). Further, $\{V^{[l]}_i, V^{[l]}_j\}$ is a qualified edge, so by (\ref{eq:p1}), $V^{[l]}_i - V^{[l]}_j$ can recover $W_l$ and correctness is guaranteed. To verify security, note that $(V_i, V_j)$ is invertible to $(V_i-V_j, V_i)$, which is further invertible to $(W_l, V_i)$. From (\ref{eq:c1}), (\ref{eq:c2}), $V_i$ is fully covered by uniform noise variables such that nothing about source symbols other than $W_l$ is revealed and security follows. Finally, any unqualified edge $\{V_i, V_j\}$ is easily seen to be secure, because if $V_i, V_j$ belong to the same qualified component, then $V_i = V_j$ and $V_i$ is independent of all source symbols; otherwise $V_i, V_j$ belong to two qualified components, then $V_i, V_j$ are covered by independent noise variables.

\section{Proof of Theorem \ref{thm:d2}} \label{sec:thm2}
This section contains the proof of Theorem \ref{thm:d2}. We first prove the `only if' part in Section \ref{sec:thm21} and then prove the `if' part in Section \ref{sec:thm22}.

\subsection{Only if Part}\label{sec:thm21}
We start with a useful lemma that holds for any symbol rate and any graph. This lemma will  be used in the proof of Theorem \ref{thm:2d} as well.

\begin{lemma}[Independence of Non-common Sources] \label{lemma:ind} 
A coded symbol $V$ must be independent of its non-common source symbols (with and without conditioning on the common source symbols),
\begin{eqnarray}
&& I\left(V; \left( W_k \right)_{k\in [K]\backslash \mathcal{C}(V)} \mid \left(W_k\right)_{k\in\mathcal{C}(V)} \right) = 0, \label{eq:ind} \\ 
&& I\left(V; \left( W_k \right)_{k\in [K]\backslash \mathcal{C}(V)} \right) = 0. \label{eq:ind0} 
\end{eqnarray}
\end{lemma}

{\it Proof:} First, we prove (\ref{eq:ind}). Consider any non-common source symbol $W_i$ of the node $V$, i.e., $i \in [K]\backslash \mathcal{C}(V)$. As $W_i$ is not a common source symbol of $V$, there must exist an edge $\{V, V_j\}$ such that $i \notin t(\{V, V_j\}) = \mathcal{D}$, for which from the security constraint (\ref{sec}) we have
\begin{eqnarray}
0 &\overset{(\ref{sec})}{=}& I\left(V, V_j; \left(W_k\right)_{k \in [K]\backslash\mathcal{D}} \mid \left(W_k\right)_{k \in \mathcal{D}} \right) \\
&\geq& I\left(V; W_i \mid \left(W_k\right)_{k \in [K]\backslash\{i\} } \right). \label{eq:ind1}
\end{eqnarray}
Consider any subset $\mathcal{J}$ of $[K]\backslash(\{i\}\cup \mathcal{C}(V))$. 
As the source symbols $W_k$ are independent (refer to (\ref{h1})), from (\ref{eq:ind1}) we have
\begin{eqnarray}
0 &\overset{(\ref{eq:ind1})}{\geq}& I\left(V; W_i \mid \left(W_k\right)_{k \in [K]\backslash\{i\} } \right) \\
&\overset{(\ref{h1})}{=}& I\left(V, \left(W_k\right)_{k \in [K]\backslash( \{i\} \cup \mathcal{C}(V) \cup \mathcal{J}) } ; W_i \mid \left(W_k\right)_{k \in \mathcal{J}}, \left(W_k\right)_{k \in \mathcal{C}(V)} \right) \\
&\geq& I\left(V; W_i \mid \left(W_k\right)_{k \in \mathcal{J}}, \left(W_k\right)_{k \in \mathcal{C}(V)} \right). \label{eq:ind2}
\end{eqnarray}
The desired identity (\ref{eq:ind}) can now be obtained by adding (\ref{eq:ind2}) for a proper sequence of $\mathcal{J}$ that is consistent with the chain rule expansion of (\ref{eq:ind}).

Second, we prove (\ref{eq:ind0}), as a simple consequence of (\ref{eq:ind}).
\begin{eqnarray}
0 &\overset{(\ref{eq:ind})}{=}& I\left(V; \left( W_k \right)_{k\in [K]\backslash \mathcal{C}(V)} \mid \left(W_k\right)_{k\in\mathcal{C}(V)} \right) \\
&\overset{(\ref{h1})}{=}& I\left(V, \left(W_k\right)_{k\in\mathcal{C}(V)}; \left( W_k \right)_{k\in [K]\backslash \mathcal{C}(V)}  \right)\\
&\geq& I\left(V; \left( W_k \right)_{k\in [K]\backslash \mathcal{C}(V)}  \right).
\end{eqnarray}
\hfill \QED

We now proceed to the proof of the `only if' part.
We show that symbol rate $R = 1/D$ is not achievable if a graph $G$ does not satisfy the condition in Theorem \ref{thm:d2}, i.e., there exists a qualified component $Q$ of the non-degenerate subgraph $\widetilde{G}$ of $G$ such that the characteristic graph $Q^{[k]}$ of some coded symbol $W_k$ contains an internal qualified edge. Without loss of generality, suppose $k = 1$ and suppose the internal qualified edge is $\{V_1^{[1]}, V_P^{[1]}\}$, which is inside the sequence of unqualified edges $\left(\{V_1^{[1]}, V_2^{[1]}\}, \cdots, \{V_{P-1}^{[1]}, V_P^{[1]}\}\right)$.
To set up the proof by contradiction, let us assume that $R = \frac{1}{D} =  \lim_{L_w\rightarrow \infty}\frac{L_w}{L_v}$ is asymptotically achievable, i.e., ${\color{black}L_v=DL_w+o(L_w)}$. Note that when the rate is exactly achievable, the $o(L_w)$ term is  zero and the following proof continues to hold.

We show that when $R=1/D$, each non-degenerate coded symbol that has no common source must be fully covered by noise. This property is stated in the following lemma.

\begin{lemma}[Noise Size] \label{lemma:size}
When $R = 1/D$, for a non-degenerate graph $\widetilde{G} = (\mathcal{V}, \mathcal{E})$ such that every node $V \in \mathcal{V}$ satisfies $\mathcal{C}(V) = \emptyset$, we have 
\begin{eqnarray}
H(V) = H\left(V \mid \left(W_k\right)_{k \in [K]\backslash\{1\}}\right) = H\left(V \mid \left(W_k\right)_{k \in [K]}\right) = DL_w {\color{black}+o(L_w)}. \label{eq:size}
\end{eqnarray}
\end{lemma}

{\it Proof:} 
From (\ref{eq:ind0}) and $\mathcal{C}(V) = \emptyset$, we have $H(V) = H\left(V \mid \left(W_k\right)_{k \in [K]\backslash\{1\}}\right) = H\left(V \mid \left(W_k\right)_{k \in [K]}\right)$. Noting that $H(V) \leq L_v = DL_w{\color{black}+o(L_w)}$, we only need to prove $H(V) \geq DL_w{\color{black}+o(L_w)}$ and this is presented next.

As $V$ is non-degenerate, there exists a qualified edge $\{V,V_i\} \in \mathcal{E}$ and $t(\{V, V_i\}) = \mathcal{D}, |\mathcal{D}| = D$. From the correctness constraint (\ref{dec}), we have
\begin{eqnarray}
DL_w &\overset{(\ref{h1})}{=}& H\left( \left(W_k \right)_{k\in\mathcal{D}} \right)\\
&\overset{(\ref{dec})}{=}& I\left(V, V_i; \left(W_k \right)_{k\in\mathcal{D}} \right) \\
&\overset{(\ref{eq:ind0})}{=}& I\left(V; \left(W_k \right)_{k\in\mathcal{D}} \mid V_i \right) \label{eq:s11}\\
&\leq& H(V)
\end{eqnarray}
where in (\ref{eq:s11}), we use the fact that $V_i \in \mathcal{V}\backslash\mathcal{V}_d$ such that $\mathcal{C}(V_i) = \emptyset$ and from Lemma \ref{lemma:ind}, $V_i$ is independent of the source symbols $W_k$. 
\hfill\QED

Next we show that when $R=1/D$, all nodes in a qualified component must use the same noise, i.e., noise must align. This property is stated in the following lemma.

\begin{lemma}[Noise Alignment] \label{lemma:noise}
When $R = 1/D$, for a non-degenerate graph $\widetilde{G} = (\mathcal{V}, \mathcal{E})$ such that every node $V \in \mathcal{V}$ satisfies $\mathcal{C}(V) = \emptyset$, we have 
\begin{eqnarray}
&& \forall \{V_i, V_j\}\in\mathcal{E} ~\mbox{such that}~t(\{V_i,V_j\})=\mathcal{D}, |\mathcal{D}|=D, ~H\left(V_i, V_j \mid \left(W_k\right)_{k \in [K]} \right) = DL_w{\color{black}+o(L_w)},\notag\\ &&\label{eq:noise} \\
&& \mbox{and for any qualified component $Q$ of $\widetilde{G}$ with node index set $\mathcal{Q}$},\notag\\
&& ~H\left((V_i)_{i\in\mathcal{Q}} \mid \left(W_k\right)_{k \in [K]} \right) = DL_w{\color{black}+o(L_w)}. \label{eq:noise0}
\end{eqnarray}
\end{lemma}

{\it Proof:} First, consider (\ref{eq:noise}). On the one hand, we have
\begin{eqnarray}
H\left(V_i, V_j \mid \left(W_k\right)_{k \in [K]} \right) &=& H\left(V_i, V_j \right) - I\left(V_i, V_j; \left(W_k\right)_{k \in [K]} \right) \\
&\overset{(\ref{dec})}{=}& H\left(V_i, V_j \right) - I\left(V_i, V_j, \left(W_k\right)_{k \in \mathcal{D}}; \left(W_k\right)_{k \in [K]} \right) \\
&\leq& H\left(V_i, V_j \right) - H\left( \left(W_k\right)_{k \in \mathcal{D}}\right) \\
&\overset{(\ref{h1})}{\leq}& 2L_v - DL_w = DL_w{\color{black}+o(L_w)}.
\end{eqnarray}
On the other hand, we have
\begin{eqnarray}
H\left(V_i, V_j \mid \left(W_k\right)_{k \in [K]} \right) \geq H\left(V_i \mid \left(W_k\right)_{k \in [K]} \right) \overset{(\ref{eq:size})}{=} DL_w{\color{black}+o(L_w)}.
\end{eqnarray}

Second, consider (\ref{eq:noise0}). 
The ``$\geq$'' direction is obvious, because for any $j \in \mathcal{Q}$
\begin{eqnarray}
H\left((V_i)_{i\in\mathcal{Q}} \mid \left(W_k\right)_{k \in [K]} \right)  \geq H\left(V_j \mid \left(W_k\right)_{k \in [K]} \right) \overset{(\ref{eq:size})}{=} DL_w{\color{black}+o(L_w)}
\end{eqnarray}
and we only need to prove the ``$\leq$'' direction. Start with any qualified edge $\{V_{i_1}, V_{i_2}\}, i_1, i_2 \in \mathcal{Q}$ in the qualified component $Q$, inside which there must exist a node $V_{i_3}$ and a node from $V_{i_1}, V_{i_2}$ (suppose it is $V_{i_2}$ without loss of generality) such that $\{V_{i_2}, V_{i_3}\}$ is a qualified edge. From the sub-modularity property of entropy functions, we have
\begin{eqnarray}
&& H\left(V_{i_1}, V_{i_2} \mid \left(W_k\right)_{k \in [K]}\right) + H\left(V_{i_2}, V_{i_3} \mid \left(W_k\right)_{k \in [K]}\right) \notag\\
&\geq& H\left(V_{i_1}, V_{i_2}, V_{i_3} \mid \left(W_k\right)_{k \in [K]}\right) + H\left( V_{i_2} \mid \left(W_k\right)_{k \in [K]}\right) \\
\overset{(\ref{eq:noise}) (\ref{eq:size})}{\Longrightarrow} && DL_w + DL_w \geq H\left(V_{i_1}, V_{i_2}, V_{i_3} \mid \left(W_k\right)_{k \in [K]}\right) + DL_w{\color{black}+o(L_w)}\\
\Rightarrow ~~~&& H\left(V_{i_1}, V_{i_2}, V_{i_3} \mid \left(W_k\right)_{k \in [K]}\right) \leq DL_w{\color{black}+o(L_w)}.
\end{eqnarray}
Then we can similarly proceed to include all nodes in $Q$. As $Q$ is a qualified component, there must exist a vertex $V_{i_4}, i_4 \in \mathcal{Q}$ such that $\{V, V_{i_4}\}$ is a qualified edge, where $V$ is one vertex from $V_{i_1}, V_{i_2}, V_{i_3}$. Similarly, we have
\begin{eqnarray}
H\left(V_{i_1}, V_{i_2}, V_{i_3}, V_{i_4} \mid \left(W_k\right)_{k \in [K]}\right) &\leq& DL_w{\color{black}+o(L_w)}, ~~\cdots, \notag\\
H\left((V_i)_{i\in\mathcal{Q}} \mid \left(W_k\right)_{k \in [K]} \right) &\leq& DL_w{\color{black}+o(L_w)}.
\end{eqnarray}
\hfill\QED

Consider the nodes $V_1, \cdots, V_P$ that violate the condition in Theorem \ref{thm:d2}, i.e., from each one of $(V_1, V_2)$, $\cdots$,  $(V_{P-1}, V_P)$, we cannot learn anything about $W_1$; from $(V_1, V_P)$, we can decode $W_1$. In the following lemma, we show that the coded symbols $V_1, \cdots, V_P$ must contain the same information of $W_1$ and noise, i.e., the coded symbols must align.

\begin{lemma}[Coded Symbol Alignment] \label{lemma:signal}
When $R = 1/D$, for the nodes $V_1, \cdots, V_P$ as specified above, we have
\begin{eqnarray}
\forall p\in[P-1], && H\left(V_{p}, V_{p+1} \mid \left(W_k\right)_{k\in[K]\backslash\{1\}} \right) = DL_w{\color{black}+o(L_w)}, \label{eq:signal}\\
&& H\left(V_1, V_P \mid \left(W_k\right)_{k\in[K]\backslash\{1\}} \right) = DL_w{\color{black}+o(L_w)}. \label{eq:signal0}
\end{eqnarray}
\end{lemma}

{\it Proof:} For both (\ref{eq:signal}) and (\ref{eq:signal0}), the ``$\geq$'' direction follows from (\ref{eq:size}) and we only need to prove the ``$\leq$'' direction.

First, consider (\ref{eq:signal}). From $(V_p, V_{p+1})$ we cannot decode $W_1$, so the edge $\{V_p, V_{p+1}\}$ is either an unqualified edge or a qualified edge but $1 \notin t(\{V_p, V_{p+1}\})$. In the former case, from the security constraint (\ref{sec}) where $t(\{V_p, V_{p+1}\}) = \emptyset$, we have
\begin{eqnarray}
 H\left(V_{p}, V_{p+1} \mid \left(W_k\right)_{k\in[K]\backslash\{1\}} \right)  &\overset{(\ref{sec})(\ref{h1})}{=}&  H\left(V_{p}, V_{p+1} \mid \left(W_k\right)_{k\in[K]} \right) \\
 &\leq&  H\left( \left(V_i\right)_{i\in\mathcal{Q}} \mid \left(W_k\right)_{k\in[K]} \right) \label{eq:s2}\\
 &\overset{(\ref{eq:noise0})}{=}& DL_w{\color{black}+o(L_w)}
\end{eqnarray}
where (\ref{eq:s2}) follows from the fact that $V_1,\cdots,V_P$ belong to a qualified component $Q$ with node index set $\mathcal{Q}$.
In the latter case, from the correctness constraint (\ref{dec}) where $1\notin t(\{V_p, V_{p+1}\}) = \mathcal{D}$, we have
\begin{eqnarray}
H\left(V_{p}, V_{p+1} \mid \left(W_k\right)_{k\in[K]\backslash\{1\}} \right) &\overset{(\ref{dec})}{=}& H\left(V_{p}, V_{p+1}\right) - I\left(V_{p}, V_{p+1},\left(W_k\right)_{k \in \mathcal{D}} ; \left(W_k\right)_{k\in[K]\backslash\{1\}} \right) \\
&\overset{(\ref{h1})}{\leq}& 2L_v - DL_w = DL_w{\color{black}+o(L_w)}.
\end{eqnarray}

Second, consider (\ref{eq:signal0}). From the sub-modularity property of entropy functions, we have
\begin{eqnarray}
(P-1)DL_w + o(L_w) &\overset{(\ref{eq:signal})}{=}& \sum_{p\in[P-1]} H\left(V_{p}, V_{p+1} \mid \left(W_k\right)_{k\in[K]\backslash\{1\}} \right)\\
&\geq& H\left(V_1, \cdots, V_P \mid \left(W_k\right)_{k\in[K]\backslash\{1\}} \right) + \sum_{p=2}^{P-1} H\left(V_p \mid \left(W_k\right)_{k\in[K]\backslash\{1\}} \right) \\
&\overset{(\ref{eq:size})}{\geq}& H\left(V_1, V_P \mid \left(W_k\right)_{k\in[K]\backslash\{1\}} \right) + (P-2) DL_w{\color{black}+o(L_w)}\\
\Rightarrow && H\left(V_1, V_P \mid \left(W_k\right)_{k\in[K]\backslash\{1\}} \right) \leq DL_w{\color{black}+o(L_w)}.
\end{eqnarray}
\hfill \QED

After establishing the above lemmas, we are ready to demonstrate the contradiction as follows. Recall that from $(V_1,V_P)$, we can recover $W_1$, i.e., $1\in t(\{V_1,V_P\})$. 
\begin{eqnarray}
DL_w{\color{black}+o(L_w)} &\overset{(\ref{eq:signal0})}{=}& H\left(V_1, V_P \mid \left(W_k\right)_{k\in[K]\backslash\{1\}} \right) \\
&\overset{(\ref{dec})}{=}& H\left(V_1, V_P, W_1 \mid \left(W_k\right)_{k\in[K]\backslash\{1\}} \right) \\
&=& H\left( W_1 \mid \left(W_k\right)_{k\in[K]\backslash\{1\}} \right) + H\left(V_1, V_P \mid \left(W_k\right)_{k\in[K]} \right) \\
&\overset{(\ref{h1})(\ref{eq:noise})}{=}&  L_w + DL_w{\color{black}+o(L_w)}. \label{eq:d1}
\end{eqnarray}
{\color{black}Normalizing (\ref{eq:d1}) by $L_w$ and letting $L_w$ approach infinity, we have $D = 1+D$, and the contradiction is arrived. The proof of the only if part is thus complete.}

\subsection{If Part}\label{sec:thm22}
We show that if the condition in Theorem \ref{thm:d2} is satisfied, then the secure storage capacity is $1/D$. We first prove that $R \leq 1/D$ and then show that $R = 1/D$ is achievable.

The proof of $R \leq 1/D$ is immediate. As $\widetilde{G}$ is not empty, there must exist a qualified edge $\{V_i, V_j\}$ such that $t(\{V_i, V_j\}) = \mathcal{D}, |\mathcal{D}| = D$. From the correctness constraint (\ref{dec}), we have 
\begin{eqnarray}
DL_w &\overset{(\ref{h1})}{=}& H\left(\left(W_k\right)_{k\in\mathcal{D}}\right)\\
&\overset{(\ref{dec})}{=}& I\left(V_i, V_j; \left(W_k\right)_{k\in\mathcal{D}}\right)\\
&\overset{(\ref{eq:ind0})}{=}& I\left(V_j; \left(W_k\right)_{k\in\mathcal{D}} | V_i \right) \label{eq:s1}\\
&\leq& H(V_j) ~\leq~ L_v \\
\Rightarrow ~~ R &\overset{(\ref{rate})}{=}& L_w/L_v ~\leq~ 1/D
\end{eqnarray}
where (\ref{eq:s1}) follows from the condition that $\mathcal{C}(V_i) = \emptyset, \forall V_i \in \mathcal{V}\backslash\mathcal{V}_d$ and (\ref{eq:ind0}).

We now present a secure storage code construction that achieves symbol rate $R = 1/D$ if $G = (\mathcal{V}, \mathcal{E})$ satisfies the condition in Theorem \ref{thm:d2}. The scheme is a generalization of that presented in Section \ref{sec:thm11}.
Suppose $L_w = \log_2(q)$ bits and each source symbol $W_k$ is one symbol from finite field $\mathbb{F}_q$, where $q>D|\mathcal{E}|$. Each coded symbol $V_n$ will be set as $D$ symbols from $\mathbb{F}_q$, i.e., $L_v = D \log_2(q)$ bits and $R = L_w/L_v = 1/D$, as desired.

Degenerate nodes $\mathcal{V}_d$ (and their connected edges) are trivial and we only need to consider the non-degenerate subgraph $\widetilde{G}$ of $G$. Suppose $\widetilde{G}$ has $M$ qualified components, $Q_1, \cdots, Q_M$. 

{\bf Consider $Q_m^{[k]}, \forall m \in [M], k \in [K]$.} Suppose $Q_m^{[k]}$ contains $U_m^{[k]}$ unqualified components. 
\begin{eqnarray}
&& \mbox{For each node $V^{[k]}$ in the $u$-th unqualified component of $Q_m^{[k]}$ where $u\in[U_m^{[k]}]$,}\notag\\
&& ~~\mbox{set $V^{[k]} = {\bf h}_{m,u}^{[k]} \times W_k +  {\bf Z}_m^{[k]}$} \label{eq:cc1}
\end{eqnarray}
where ${\bf h}_{m,u}^{[k]}\in\mathbb{F}_q^{D\times 1}$; ${\bf Z}_m^{[k]} \in \mathbb{F}_{q}^{D\times 1}, \forall m, k$ are i.i.d. uniform noise symbols that are independent of $W_k$. 

{\bf Consider $Q_m, \forall m \in [M]$.}
\begin{eqnarray}
\mbox{For each node $V$ in $Q_m$, set $V = \sum_{k\in[K]} V^{[k]}$.} \label{eq:cc2}
\end{eqnarray}

{\bf We show that there exists a choice of ${\bf h}_{m,u}^{[k]}, k\in[K], m\in[M], u\in[U_m^{[k]}]$ such that the code construction (\ref{eq:cc1}), (\ref{eq:cc2}) is correct and secure.} To this end, choose every entry of ${\bf h}_{m,u}^{[k]}$ independently and uniformly from $\mathbb{F}_q$. Consider correctness.
For any qualified edge $\{V_i, V_j\}$, i.e., $t(\{V_i,V_j\}) = \mathcal{D}, |\mathcal{D}| = D$, we have
\begin{eqnarray}
V_i - V_j &\overset{(\ref{eq:cc2})}{=}& \sum_{k\in[K]} \left( V^{[k]}_i - V^{[k]}_j \right) \\
&\overset{(\ref{eq:cc1})}{=}& \sum_{k\in\mathcal{D}} \left( V^{[k]}_i - V^{[k]}_j \right) \label{eq:t1}\\
&=& {\bf H}_{ij} \times \left(W_k\right)_{k\in\mathcal{D}} \label{eq:t2}
\end{eqnarray}
where (\ref{eq:t1}) from the fact that $\{V^{[k]}_i, V^{[k]}_j\}, k \notin \mathcal{D}$ is an unqualified edge such that $V^{[k]}_i, V^{[k]}_j$ belong to the same unqualified component and from (\ref{eq:cc1}), $V^{[k]}_i = V^{[k]}_j, k \notin \mathcal{D}$. (\ref{eq:t2}) is obtained because $\{V^{[k]}_i, V^{[k]}_j\}, k \in \mathcal{D}$ is a qualified edge that is not internal, i.e., spans different unqualified components. In addition, ${\bf H}_{ij}$ is a $D\times D$ matrix over $\mathbb{F}_q$, whose entries can be obtained from ${\bf h}_{m,u}^{[k]}$. View the determinant of ${\bf H}_{ij}$, $|{\bf H}_{ij}|$ as a polynomial in variables ${\bf h}_{m,u}^{[k]}, k\in[K], m\in[M], u\in[U_m^{[k]}]$. This determinant polynomial has degree $D$ and is not a zero polynomial as there exists a realization of ${\bf h}_{m,u}^{[k]}$ such that the determinant is not zero. Consider the product of the determinant polynomials for all $|\mathcal{E}|$ qualified edges,
\begin{eqnarray}
\mbox{poly}~\triangleq \prod_{i,j : \{V_i, V_j\}\in \mathcal{E}} |{\bf H}_{ij}|
\end{eqnarray}
which is a non-zero polynomial and has degree at most $D|\mathcal{E}|$. By the Schwartz–Zippel lemma \cite{Demillo_Lipton, Schwartz, Zippel}, a uniform choice of ${\bf h}_{m,u}^{[k]}, k\in[K], m\in[M], u\in[U_m^{[k]}]$ over $\mathbb{F}_q$ where $q > D|\mathcal{E}|$ (the degree of $\mbox{poly}$) guarantees $\mbox{poly}$ is not always zero. It follows that there exists some realization of ${\bf h}_{m,u}^{[k]}, k\in[K], m\in[M], u\in[U_m^{[k]}]$ such that $\mbox{poly} \neq 0$. Then each $|{\bf H}_{ij}|$ is not zero and from each qualified edge, we can recover the $D$ desired source symbols, i.e., correctness is guaranteed.

Finally consider security. For any qualified edge $\{V_i,V_j\}$, security is guaranteed by noting that $(V_i, V_j)$ is invertible to $\left(\left(W_k\right)_{k\in\mathcal{D}}, V_i\right)$ and $V_i$ is fully covered by uniform noise variables. For any unqualified edge $\{V_i, V_j\}$, security holds no matter whether $V_i, V_j$ belong to the same qualified component (same coded symbol assignment, i.e., $V_i=V_j$) or two qualified components (then $V_i, V_j$ are protected by independent noise variables).


\section{Proof of Theorem \ref{thm:2d}} \label{sec:thm3}
This section contains the proof of Theorem \ref{thm:2d}. 
We first prove the `only if' part in Section \ref{sec:thm31} and then prove the `if' part in Section \ref{sec:thm32}.

\subsection{Only if Part}\label{sec:thm31}
We start with a useful property for any secure storage code of symbol rate $R = 2/D$, stated in the following lemma. Note that when $R = \frac{2}{D} = \lim_{L_w \rightarrow \infty} \frac{L_w}{L_v}$, we have\footnote{The same proof holds when the $o(L_w)$ term is 0, i.e., when the rate is exactly achievable.}
\begin{eqnarray}
2L_v = DL_w + o(L_w). \label{arate}
\end{eqnarray}

\begin{lemma}[Deterministic of Common Sources] \label{lemma:det}
When $R = 2/D$, a coded symbol $V$ that is connected to a qualified edge is asymptotically deterministic given its common source symbols,
\begin{eqnarray}
H\left(V \mid \left(W_k \right)_{k \in \mathcal{C}(V)} \right) = o(L_w). \label{eq:det}
\end{eqnarray}
\end{lemma}

{\it Proof:} Consider any qualified edge $\{V, V_i\}$ such that $t(\{V, V_i\}) = \mathcal{D}, |\mathcal{D}| = D$. From the correctness constraint (\ref{dec}), we have
\begin{eqnarray}
2L_v &\geq& H(V, V_i) \\
&\overset{(\ref{dec})}{=}& H\left( V, V_i, \left(W_k\right)_{k \in \mathcal{D}} \right)\\
&=& H\left( \left(W_k\right)_{k \in \mathcal{D}} \right) + H\left( V, V_i \mid \left(W_k\right)_{k \in \mathcal{D}} \right) \\
&\overset{(\ref{h1})}{\geq}& D L_w + H\left( V \mid \left(W_k\right)_{k \in \mathcal{D}} \right) \\
&\geq& D L_w \\
&\overset{(\ref{arate})}{=}& 2 L_v + o(L_w).
\end{eqnarray}
The above sequence of inequalities starts and ends both with $2L_v$ (ignoring $o(L_w)$ terms), then all the inequalities must be equalities within the distortion of $o(L_w)$. In particular,
\begin{eqnarray}
H(V, V_i) = 2L_v + o(L_w),~~ H(V) = L_v + o(L_w) \label{eq:hv}
\end{eqnarray}
and
\begin{eqnarray}
o(L_w) &=& H\left( V \mid \left(W_k\right)_{k \in \mathcal{D}} \right) \\
&=& H\left( V \mid \left(W_k\right)_{k \in \mathcal{C}(V)}, \left(W_k\right)_{k \in \mathcal{D}\backslash\mathcal{C}(V)} \right) \\
&=& H\left( V \mid \left(W_k\right)_{k \in \mathcal{C}(V)} \right) - I\left(V; \left(W_k\right)_{k \in \mathcal{D}\backslash\mathcal{C}(V)} \mid \left(W_k\right)_{k \in \mathcal{C}(V)}    \right) \\
&\geq& H\left( V \mid \left(W_k\right)_{k \in \mathcal{C}(V)} \right) - I\left(V; \left(W_k\right)_{k \in [K]\backslash\mathcal{C}(V)} \mid \left(W_k\right)_{k \in \mathcal{C}(V)} \right) \\
&\overset{(\ref{eq:ind})}{=}& H\left( V \mid \left(W_k\right)_{k \in \mathcal{C}(V)} \right).
\end{eqnarray}
\hfill \QED

Equipped with Lemma \ref{lemma:det}, we are ready to present the proof of the `only if' part. We show that if either of the two conditions in Theorem \ref{thm:2d} is violated, then the symbol rate $R$ cannot be $2/D$. We will prove this by contradiction, so suppose there exists a secure storage code of symbol rate $R = 2/D$.

{\bf Suppose condition 1 is violated}, i.e., there exists a node $V$ such that $|\mathcal{C}(V)| < D/2$. Then
\begin{eqnarray}
L_v + o(L_w) &\overset{(\ref{eq:hv})}{=}& H(V) \\
&=& \underbrace{H\left(V \mid \left(W_k \right)_{k \in \mathcal{C}(V)} \right)}_{\overset{(\ref{eq:det})}{=}~o(L_w)} + I\left(V; \left(W_k \right)_{k \in \mathcal{C}(V)} \right) \\
&\leq& H\left( \left(W_k\right)_{k \in \mathcal{C}(V)} \right) + o(L_w) \overset{(\ref{h1})}{=} |\mathcal{C}(V)| \times L_w + o(L_w)\\
&<& D/2 \times L_w + o(L_w) \\
\Rightarrow ~~ R &=& \lim_{L_w\rightarrow \infty} L_w/L_v ~>~ 2/D
\end{eqnarray}
which contradicts the assumption that $R = 2/D$.

{\bf Suppose condition 1 is satisfied while condition 2 is violated}, i.e., there exists a qualified edge $\{V_i, V_j\}$ such that $\mathcal{C}(V_i)\cup \mathcal{C}(V_j)$ is a strict subset of $t(\{V_i, V_j\})$. Then $|\mathcal{C}(V_i)\cup \mathcal{C}(V_j)| < D$ and
\begin{eqnarray}
2L_v + o(L_w) &\overset{(\ref{eq:hv})}{=}& H(V_i, V_j) \\
&=& \underbrace{H\left(V_i, V_j \mid \left(W_k \right)_{k \in \mathcal{C}(V_i) \cup \mathcal{C}(V_j)} \right)}_{\overset{(\ref{eq:det})}{=}~o(L_w)} + I\left(V_i, V_j ; \left(W_k \right)_{k \in \mathcal{C}(V_i)\cup \mathcal{C}(V_j)} \right)\\
&\leq& H\left( \left(W_k \right)_{k \in \mathcal{C}(V_i) \cup \mathcal{C}(V_j)} \right) +o(L_w) \overset{(\ref{h1})}{=} |\mathcal{C}(V_i)\cup \mathcal{C}(V_j)| \times L_w + o(L_w) \\
&<& D \times L_w + o(L_w)\\
\Rightarrow ~~ R &=& \lim_{L_w\rightarrow \infty} L_w/L_v ~>~ 2/D
\end{eqnarray}
which contradicts the assumption that $R = 2/D$.

\subsection{If Part}\label{sec:thm32}
We show that if the two conditions in Theorem \ref{thm:2d} are satisfied, then the secure storage capacity is $2/D$. 
We first prove that $R \leq 2/D$ and then show that $R = 2/D$ is achievable.

The proof of $R \leq 2/D$ is immediate. As condition 1 is satisfied, all edges are qualified. Pick any one, say $\{V_i, V_j\}$ such that $t(\{V_i, V_j\}) = \mathcal{D}, |\mathcal{D}| = D$. From the correctness constraint (\ref{dec}), we have
\begin{eqnarray}
2L_v &\geq& H(V_i, V_j) \\
&\overset{(\ref{dec})}{=}& H\left(V_i, V_j, \left(W_k\right)_{k\in\mathcal{D}} \right) \\
&\geq& H\left(\left(W_k\right)_{k\in\mathcal{D}} \right) \\
&\overset{(\ref{h1})}{=}& D \times L_w \\
\Rightarrow ~~ R &\overset{(\ref{rate})}{=}& L_w/L_v ~\leq~ 2/D.
\end{eqnarray}

We now present a secure storage code construction that achieves symbol rate $R = 2/D$. Consider any graph $G = (\mathcal{V}, \mathcal{E})$ that satisfies the two conditions in Theorem \ref{thm:2d}. Set $L_w = 2 \log_2(q)$ bits, where $q > 2D|\mathcal{E}|$. Suppose each $W_k$ consists of $2$ i.i.d. uniform symbols from $\mathbb{F}_q$, i.e., $W_k \in \mathbb{F}_q^{2\times 1}$. We set each coded symbol $V_1, \cdots, V_N$ as follows so that $V_n \in \mathbb{F}_q^{D\times 1}, \forall n \in [N]$, i.e., $L_v = D \log_2(q)$ bits and the symbol rate achieved is $R = L_w/L_v = 2/D$, as desired.
\begin{eqnarray}
\mbox{Set}~V_n = {\bf H}_{n} \times \left(W_k\right)_{k \in \mathcal{C}(V_n)}, \forall n \in [N] \label{eq:cc}
\end{eqnarray}
where $\left(W_k\right)_{k \in \mathcal{C}(V_n)} \in \mathbb{F}_q^{2|\mathcal{C}(V_n)| \times 1}$ is a column vector that stacks each $W_k$ and ${\bf H}_{n} \in \mathbb{F}_q^{D\times 2|\mathcal{C}(V_n)|}$.

Next we show that there exists a choice of ${\bf H}_n, n \in [N]$ so that the constructed code satisfies the correctness and security constraints (\ref{dec}), (\ref{sec}). To prove the existence, we generate ${\bf H}_n, n \in [N]$ randomly by choosing each element of ${\bf H}_n, n \in [N]$ independently and uniformly from $\mathbb{F}_q$.

Note that condition 2 in Theorem \ref{thm:2d} is satisfied, i.e., for any qualified edge $\{V_i, V_j\}$ such that $t(\{V_i, V_j\}) = \mathcal{D}, |\mathcal{D}| = D$, we have $\mathcal{C}(V_i) \cup \mathcal{C}(V_j) = \mathcal{D}$.
Then from the code construction (\ref{eq:cc}), the coded symbols $(V_i, V_j)$ do not contain any undesired source symbols $\left(W_k\right)_{k \in [K]\backslash\mathcal{D}}$ so that nothing is revealed about the undesired source symbols (note that the source symbols are independent) and security is guaranteed.
Regarding correctness, for any qualified edge, from the coded symbols $(V_i, V_j)$ we have $2D$ linear combinations in the $2D$ desired source symbols. That is, the row stack of $V_i,V_j$ produces 
\begin{eqnarray}
[V_i; V_j] = {\bf H}_{ij} \times \left(W_k\right)_{k\in\mathcal{D}}
\end{eqnarray}
where ${\bf H}_{ij} \in \mathbb{F}_q^{2D \times 2D}$ can be obtained from ${\bf H}_i, {\bf H}_j, \mathcal{C}(V_i), \mathcal{C}(V_j)$. View the determinant of ${\bf H}_{ij}$, $|{\bf H}_{ij}|$ as a polynomial in variables ${\bf H}_n, n \in [N]$. This determinant polynomial has degree $2D$ and is not a zero polynomial as there exists a realization of ${\bf H}_n, n \in [N]$ such that the determinant is not zero. Consider the product of the determinant polynomials for all $|\mathcal{E}|$ qualified edges,
\begin{eqnarray}
\mbox{poly}~\triangleq \prod_{i,j : \{V_i, V_j\}\in \mathcal{E}} |{\bf H}_{ij}|
\end{eqnarray}
which is a non-zero polynomial and has degree at most $2D|\mathcal{E}|$. By the Schwartz–Zippel lemma \cite{Demillo_Lipton, Schwartz, Zippel}, a uniform choice of ${\bf H}_n, n \in [N]$ over the finite field $\mathbb{F}_q$ where $q > 2D|\mathcal{E}|$ (the degree of $\mbox{poly}$) guarantees $\mbox{poly}$ is not always zero. It follows that there exists some realization of ${\bf H}_n, n \in [N]$ such that $\mbox{poly} \neq 0$. Then each $|{\bf H}_{ij}|$ is not zero and from each qualified edge, we can recover all desired source symbols, i.e., correctness is guaranteed.

\section{Discussion}
In this work we have formulated a problem on secure storage under data access and security constraints specified by graphs and considered the maximum storage efficiency - capacity, as the performance metric. We have focused on extremal graphs where the capacity takes extremal values (e.g., maximum with non-trivial security constraints). The extremal graph characterizations obtained in this work are guided by an alignment view that is effective for both code constructions and impossibility claims. For the extremal rates considered, a crucial graphical structure turns out to be `internal qualified edges', which capture the tension between using the same noise and storing the same coded symbol for security, and diversifying the coded symbols for correctness. 

Similar to many challenging open problems in network information theory, allowing arbitrary network topologies often includes intractable problem instances. The perspective we take in this work is to concentrate on extremal networks and study the consequences of the extremal structures. While we have exclusively focuses on networks with extremal rates (and special extremal values), many other choices appear promising along this line, e.g., shortest/sparest codes under smoothness/locality constraints \cite{Sun_Jafar_LDC, Kazama_Kamatsuka_Yoshida_Matsushima} and might lead to new interesting questions and solutions.

\let\url\nolinkurl
\bibliographystyle{IEEEtran}
\bibliography{Thesis}
\end{document}